\documentclass[twocolumn]{aastex701}
\usepackage{float}

\usepackage{hyperref}
\usepackage{footmisc}
\usepackage{amsmath}

\shorttitle{Orbital Data Centers and Rubin Observatory}

\begin{document}

\title{The Impact of Orbital Data Centers on the NSF-DOE Vera C. Rubin Observatory}

\author[0009-0000-3705-3753]{Nayan Jangid}
\affiliation{The Grainger College of Engineering, Department of Aerospace Engineering, University of Illinois Urbana-Champaign, Urbana, IL, USA}
\affiliation{IAU Centre for the Protection of the Dark and Quiet Sky, France}
\email{njangid2@illinois.edu}
\author[0000-0003-1305-7308]{Meredith L. Rawls}
\affiliation{Department of Astronomy / DiRAC / Vera C. Rubin Observatory, University of Washington, Seattle, WA, USA}
\affiliation{IAU Centre for the Protection of the Dark and Quiet Sky, France}
\email{mrawls@uw.edu}
\author[0000-0003-2874-6464]{Peter Yoachim}
\affiliation{Department of Astronomy / DiRAC / Vera C. Rubin Observatory, University of Washington, Seattle, WA, USA}
\affiliation{IAU Centre for the Protection of the Dark and Quiet Sky, France}
\email{yoachim@uw.edu}
\author[0000-0003-0064-4298]{Connie Walker}
\affiliation{NSF NOIRLab, Tucson, AZ, USA}
\affiliation{IAU Centre for the Protection of the Dark and Quiet Sky, France}
\email{connie.walker@noirlab.edu}
\author[0000-0002-1398-6302]{Siegfried Eggl}
\affiliation{The Grainger College of Engineering, Department of Aerospace Engineering, University of Illinois Urbana-Champaign, Urbana, IL, USA}
\affiliation{Department of Astronomy, University of Illinois Urbana-Champaign, Urbana, IL, USA}
\affiliation{IAU Centre for the Protection of the Dark and Quiet Sky, France}
\email{eggl@illinois.edu},

\begin{abstract}
As the NSF-DOE Vera C. Rubin Observatory begins its Legacy Survey of Space and Time (LSST), satellite operators are
racing to launch Orbital Data Center (ODC) satellites into low-Earth orbit.
We present an impact assessment of a simulated $\sim$1 million ODC satellite constellation, arranged per recent filings across two low-Earth and two Sun-synchronous orbit regimes, on one year of Rubin Observatory LSST observations. Our six-stage sequential pipeline combines orbit propagation, field-of-view matching, observer-frame geometry, brightness modeling, surface-brightness calculations, and pixel-loss estimates.
We find that on a given night, up to 80\% of LSST pointings would contain at least one illuminated ODC satellite streak, depending on the time of year. The yearly average of contaminated exposures is closer to 20\%, due to seasonal visibility variations and bright satellite streaks being most prominent near twilight.
Sun-synchronous ODCs disproportionately contaminate due to preferential illumination geometry.
The ODC satellite brightness distribution is multi-modal, where solar panel specular reflection can appear as bright as $2-4$ AB mag ($\mu_g \approx 9-11\,$ mag arcsec$^{-2}$), while diffuse reflections more typically reach $6$ AB mag ($\mu_g \approx 13-14\,$ mag arcsec$^{-2}$).
Even at 10\% deployment, the constellation would still produce widespread visible streak contamination throughout the year.
ODC satellites would also drastically increase sky brightness --- by more than 1000\% near twilight with no Moon.
Launching and operating ODC satellites at this scale would significantly harm LSST science, including typical $5-10\%$ nightly mean pixel losses near twilight, even if multiple mitigation strategies are implemented.

\end{abstract}

\keywords{Ground-based astronomy, Light pollution, Sky surveys, Artificial satellites}

\section{Introduction} \label{sec:intro}
The widespread proliferation of constellations of tens of thousands of artificial satellites in low-Earth orbit (LEO) is today a reality. Its impact on ground and space based astronomy are well documented
\citep{hainaut26} but still poorly understood in terms of science outcomes.
Satellite constellation operators have engaged the astronomy community to develop technologies that can mitigate their impact on the oldest of sciences, with some success,
but most satellites still exceed the 7th AB magnitude (a flux-calibrated photometric system with zero-point $F_0 = 3631$\,Jy) at 550 km brightness recommendation from the International Astronomical Union Centre for the Protection of the Dark and Quiet Sky \citep[IAU CPS; see][]{boley25,mallama2025satellite}.

In a similar fashion, astronomers have developed strategies to deal with some intended and unintended emissions of satellites across the electromagnetic spectrum.
The latter include operational data sharing \citep{nhan2025operational},
satellite avoidance (``satellite dodging'') \citep{hu22,dadighat2025ADASS},
satellite brightness assessment
\citep{nandakumar2023high,krantz2023steward,lorenzi2024refl,mallama2025satellite}
and modeling 
\cite{fankhauser23, jangid-in-prep}, 
as well as testing of new sensors and development of fast shutters for future astronomical facilities.
The scale of planned satellite constellations, and the increasing size of satellites, 
however, vastly outpace current mitigation efforts.

Most notably, recent filings with the US Federal Communications Commission (FCC) from SpaceX \citep{spacex_odc_2026} and Starcloud\footnote{For an up-to-date list of large constellations that have filed with the FCC or International Telecommunication Union (ITU), please see \url{https://planet4589.org/space/con/conlist.html}.}, as well as future plans from Axiom Space, Orbital, Blue Origin, Project Suncatcher, and more, have collectively proposed well over one million Orbital Data Center (ODC) satellites in LEO.
Impacts to optical astronomy from such a population will be most severe for ground- and LEO-based observatories with large fields of view and sensitive instruments designed to detect faint astrophysical sources. Other factors that affect impacts include telescope primary mirror size, telescope latitude (for ground-based) or orbital altitude (for LEO-based), and observation program and pointing strategy.

The present study assesses the impact of large constellations of ODC satellites on the NSF-DOE Vera C. Rubin Observatory Legacy Survey of Space and Time \citep[LSST,][]{lsst-main}.
This is a publicly funded international collaboration survey mission that features a large ground-based optical telescope, the Simonyi Survey Telescope, with a wide field of view, and an extremely sensitive camera, the LSST Camera.
The Rubin Project began operations in 2025, recently releasing its first Data Preview \citep{dp1}, just as the active LEO satellite population approached 15,000 in number. Most of these are commercial telecommunication constellation satellites, and at the time of this writing in 2026 June, only a few prototype ODC satellites have launched. While Rubin Observatory is certainly not the only facility affected, it will collect more data than all previous optical astronomy observations combined during the first year of LSST. Any mitigations that improve the ability of the Rubin LSST to accomplish its science goals will likely benefit all optical telescope facilities and the visible sky. We acknowledge that while impacts on other wavelength regimes, including radio, are out of scope for this study, these will also be severely affected by ODCs


In this work, we present a six-stage modeling approach to create a more complete picture of the impact of ODCs on the Rubin Observatory LSST. The stages are described in Section \ref{sec:method}. We apply the full modeling pipeline to quantify the impacts on a simulated ODC satellite constellation in Section \ref{sec:results}. In Section \ref{sec:mitigation}, we describe how changing the constellation size affects our results and consider other mitigation approaches. Finally, Section \ref{sec:science} discusses LSST science impacts and Section \ref{sec:summary} summarizes our conclusions.

\section{Methods}\label{sec:method}

We develop a multi-stage simulation pipeline to quantify how a very large ODC satellite constellation would affect observations conducted by the Rubin Observatory LSST. The goal of the pipeline is not only to determine whether satellites geometrically intersect LSST pointings, but also to identify which of those intersections represent satellites illuminated by the Sun and are therefore optically relevant, and finally to estimate the corresponding apparent brightness of the resulting streaks. In practice, these are distinct problems with different computational costs, so we treat them in a staged manner.

The pipeline consists of six sequential components, or stages: (1) identification of pointings, i.e., collections of exposures of given regions of the sky, that are potentially affected by a given orbital shell of satellites, (2) detailed propagation of satellites through flagged exposures together with field-of-view matching, (3) transformation to the observer frame and evaluation of solar illumination geometry, (4) brightness modeling for the subset of satellites that are both in the field of view and sunlit, (5) streak surface brightness computation, and (6) streak pixel loss estimation. This staged structure reduces computational cost by avoiding expensive brightness calculations for satellites that never contribute observable streaks. The pipeline code is publicly available and archived on Zenodo
\citep{jangid2026odcpipeline}.%
\footnote{\url{https://github.com/njangid2/rubin-odc-simulation-pipeline}}

\subsection{Model Orbital Data Center Constellation}\label{sec:constellation}

We base our ODC population on the proposed SpaceX Orbital Data Center system described in the FCC application by \citet{spacex_odc_2026}. For shell-by-shell orbital parameters, we use the compiled summary provided by \citet{McDowell2026}, which is based on the filing. In the full configuration, the constellation is composed of approximately one million spacecraft distributed across orbital shells spanning altitudes of 686--1002\,km. The system includes both low-inclination shells near $30^\circ$ and Sun-synchronous shells near $97^\circ$--$99.5^\circ$, producing a constellation with both broad sky coverage and a strong dependence on Sun geometry.

We model the constellation using 94 discrete orbital shells, summarized in Table~\ref{tab:shells}, for a total of 998,240 satellites. These shells are grouped into four broad regimes: a low-altitude, low-inclination component (LEO-30), a higher-altitude low-inclination component (H-LEO-30), and two Sun-synchronous orbit (SSO) components, (SSO-97 and SSO-99). The low-inclination shells are represented as 30 orbital planes with 333 satellites per plane, while the SSO shells are represented as single-plane populations with very large populations per shell, following the structure of the filing. This shell-based representation preserves the overall orbital architecture while remaining computationally tractable for survey-scale simulation.

\begin{deluxetable}{lccccr}
\tablecaption{Simulated ODC Orbital Shell Configuration\label{tab:shells}}
\tablewidth{0pt}
\tablehead{
  \colhead{Regime} &
  \colhead{Shells} &
  \colhead{Altitude\ (km)} &
  \colhead{Inclination\ ($^{\circ}$)} &
  \colhead{Planes $\times$ Sats/Plane} &
  \colhead{Total Satellites}
}
\startdata
LEO-30    & 1--25  & 686--718  & 29.3--30.5 & $30 \times 333$     & 249,750 \\
H-LEO-30  & 26--50 & 946--978  & 29.4--30.5 & $30 \times 333$     & 249,750 \\
SSO-97    & 51--72 & 707--744  & 97.1--98.2 & $1 \times 11{,}131$ & 244,882 \\
SSO-99    & 73--94 & 967--1002 & 99.4--99.5 & $1 \times 11{,}539$ & 253,858 \\
\hline
Total     & 94     & 686--1002 & \nodata    & \nodata             & 998,240 \\
\enddata
\tablecomments{Shell parameters used for a model ODC satellite constellation \citep{spacex_odc_2026,McDowell2026}.\label{tab:table1}}
\end{deluxetable}

For each shell, we generate synthetic Two-Line Element (TLE) sets at a common daily epoch. A TLE is a standardized mean-element orbital format designed for use with the SGP4 propagator and widely employed for describing Earth-orbiting satellites. Each TLE encodes the satellite's mean orbital state, including inclination, right ascension of the ascending node, eccentricity, argument of perigee, mean anomaly, mean motion, and associated epoch information. The format provides a compact and computationally efficient representation of orbital trajectories suitable for large-scale propagation studies involving extensive satellite populations.

These synthetic TLEs are not intended to reconstruct a true operational catalog, but rather to provide a self-consistent realization of the shell geometry that can be propagated with standard astrodynamics tools. In doing so, we assume circular or near-circular shell occupancy and neglect shell evolution, orbit maintenance maneuvers, launch phasing, station-keeping behavior, and deployment transients. Consequently, the generated catalog should be interpreted as an idealized steady-state realization of the proposed constellation architecture rather than a prediction of the instantaneous operational configuration. This approximation is appropriate for estimating the steady-state impact of the proposed constellation on Rubin LSST observations.

\subsection{Stage 1: Affected Exposure Identification}\label{sec:step1}

The first stage of the pipeline identifies LSST pointings that are plausibly affected by a given shell. 
The LSST employs a scheduling algorithm that uses a modified Markov Decision Process which can generate lists of desirable observations in real time \citep{naghib19}. The LSST scheduler balances the desire to minimize slew time, optimize the signal-to-noise ratio in individual images, and maintain survey footprint uniformity.
We use pointings drawn from the Rubin baseline scheduler database\footnote{\url{https://rubin-scheduler.lsst.io}\label{opsim}}, including the pointing center in right ascension and declination, exposure start time, exposure duration, filter, and observing night.
At this stage the goal is not to determine the exact number of streaks in each exposure, but rather to construct a conservative and computationally inexpensive prefilter that identifies which pointings warrant more detailed propagation.

For each shell and each calendar day, we generate a synthetic satellite population and select one representative satellite per orbital plane. These representative satellites are then propagated across the entire day using SGP4 \citep{vallado2008sgp4}.
The representative-satellite approximation is used only in this initial filtering step. Its purpose is to identify telescope pointings that are near the instantaneous sky tracks of the shell without performing the expensive computation needed to propagate all satellites in that shell.

An LSST pointing is flagged as potentially affected by a shell if any representative satellite from that shell passes within 5$^\circ$ of the pointing center at any sampled time during the day. This threshold is intentionally larger than the Simonyi field-of-view radius of 1.75$^\circ$. It therefore acts as a geometric buffer that favors recall over precision: some pointings will be carried forward even though no true crossing occurs, but pointings with genuine crossings are unlikely to be missed. In other words, this step is designed as a computational sieve rather than as the final streak classifier.

This prefiltering is essential because the full ODC population is extremely large. Directly propagating nearly one million satellites against all LSST pointings would be computationally prohibitive. In practice, Stage~1, which propagates only representative satellites to screen for plausible shell--pointing intersections, completed in approximately 11 minutes on 40 CPU cores. Stage~2, which performs full per-satellite propagation restricted to only the flagged pointing subset identified in Stage~1, still required approximately 14 hours on the same 40-core machine. Applying the full Stage~2 propagation naively to all pointings without prefiltering would therefore be substantially more expensive and computationally intractable for survey-scale simulation. By first reducing the problem to shell--day--pointing combinations that are geometrically plausible, we confine the expensive detailed propagation step to the subset of observations most likely to contain satellite streaks.

\subsection{Stage 2: Propagation and Field-of-View Matching}\label{sec:step2}

For every LSST pointing flagged in the first step, we then perform detailed propagation of all satellites belonging to that shell over the duration of the relevant exposure (typically 30 seconds). This replaces the representative-satellite approximation with a full shell realization. Satellite positions are evaluated at multiple time samples across the exposure window so that motion through the Simonyi field of view can be resolved in time rather than inferred from a single epoch. In this work, each exposure is uniformly sampled at 10 time steps, corresponding to a temporal resolution of approximately 3 seconds for a nominal 30-second Rubin exposure.

The propagated states are generated with the SGP4 orbit propagator and transformed from the True Equator Mean Equinox (TEME) reference frame into celestial right ascension and declination coordinates for comparison with LSST pointings.
From these transformed positions, we compute the right ascension and declination of each satellite at each sampled time. The angular separation between the satellite and the pointing center is then evaluated. A satellite is classified as producing a streak when its separation from the pointing center falls within the Simonyi field-of-view radius of 1.75$^\circ$. For each such crossing we record the timestamp, satellite identifier, celestial coordinates (Right Ascension and Declination), angular separation from field center, and whether the satellite lies inside the field of view at that instant.

Satellite illumination is also evaluated at this stage. For each propagated position, we determine whether the satellite is sunlit or lies within Earth's shadow using a cylindrical shadow model. The Sun's position is approximated from a low-precision ecliptic longitude formula, accurate to approximately $1^\circ$. A satellite is classified as sunlit if either it lies on the Sun-facing side of Earth, or its perpendicular distance from the Earth--Sun axis exceeds Earth's radius $R_\oplus = 6378.137$ km. This cylindrical approximation neglects the penumbra and slightly overestimates the illuminated fraction near the terminator, but is standard for survey-scale screening where precise shadow boundary geometry is not required. The illumination state is recorded alongside the geometric crossing information for each time sample.

Because this step is performed for many satellites, many exposures, and many time samples, we adopt several computational optimizations. First, time grids are constructed only for exposures previously flagged as plausible shell intersections. Second, shell processing is parallelized on a per-day basis. Third, the intermediate products are reduced to the quantities needed for downstream analysis rather than storing full propagation state vectors. These design choices preserve the physical content needed for impact assessment while keeping the workflow tractable for survey-scale simulation.

\subsection{Stage 3: Observer Geometry}\label{sec:step3}

After the celestial coordinates are computed, we transform the retained trajectories into the Rubin observer frame to obtain azimuth and elevation. This observer-frame representation is important for understanding where in the sky the visible satellites appear and for connecting the simulation output to realistic observing conditions at the Rubin site. Throughout this work, the observer location is fixed to the Vera C. Rubin Observatory site on Cerro Pachón, Chile, using an \texttt{astropy.coordinates.EarthLocation} definition with $30^\circ\,14'\,40.70''\,\mathrm{S}$,
$70^\circ\,44'\,57.90''\,\mathrm{W}$, elevation $2663\,\mathrm{m}$ above sea level.

To avoid recomputing the solar position redundantly for every shell and every satellite, the Sun azimuth and elevation at Rubin are computed once for each unique timestamp appearing in the propagated dataset and then joined onto all rows sharing that time. This is efficient because many shell trajectories share identical timestamps within the same exposure grid. The Sun elevation in particular provides a straightforward check on observing conditions: negative values indicate the Sun is below the horizon.

The output of this stage is therefore a time-resolved catalog of shell-specific satellite trajectories through LSST pointings, including celestial position, observer-frame position, field-of-view membership, and illumination state. This observer-frame transformation is the final step needed to situate satellite trajectories within the context of actual LSST observations, providing the azimuth, elevation, and solar geometry required for the brightness modeling in Section~\ref{sec:brightness}.

\subsection{Stage 4: Brightness Modeling}\label{sec:brightness}

The next stage of the pipeline estimates the apparent brightness of satellites that are both sunlit and intersect the Simonyi field of view. We compute brightness using an improved and scaled implementation based on \textsc{Lumos-Sat} \citep{fankhauser23}. In this framework, the satellite brightness is not prescribed by an assumed magnitude distribution; instead, it is derived directly from the instantaneous Sun--satellite--observer geometry and the assumed surface reflectance properties of the spacecraft. Given the satellite orientation, solar illumination, and observer viewing geometry at each time step, the reflected intensity is evaluated and subsequently converted into an apparent magnitude.

Each ODC spacecraft is modeled as a scaled-up Starlink-like platform\footnote{This is consistent with information shared by SpaceX at the time of this writing in 2026 June, i.e., \url{https://spacenews.com/spacex-offers-details-on-orbital-data-center-satellites/} and \url{https://spacenews.com/astronomers-fear-orbital-data-centers-will-interfere-with-observations/}}, with three classes of reflecting surface: the satellite bus, the solar array, and radiator panels.

For the satellite bus or chassis, we adopt a bidirectional reflectance distribution function (BRDF) based on the Starlink V2 Mini model from \citet{kandula26}, scaled to a chassis size comparable to the proposed Starlink V3 form factor of approximately $7 \times 3.5$\,m$^2$. For the solar panel, we adopt the BRDF from the Starlink V1.5 model in \citet{fankhauser23}, but scale the panel area to match a spacecraft power requirement of 400\,kW. Relative to a 25\,kW single-panel reference area of 104.96\,m$^2$, this implies a substantially larger reflective solar array and therefore a greater capacity for both diffuse and specular reflection.

To dissipate the waste heat associated with a 400\,kW-class compute payload, we additionally include a deployed radiator panel in the model. Like a flat sheet, the radiator has two opposing faces, each with an area of 110\,m$^2$, which we model as two surfaces with antiparallel normals. The radiator is modeled as a Lambertian reflector with an albedo of 0.9, reflecting a typical white or silvered thermal-control coating, which is diffuse rather than specular. In the satellite-centered frame, the $z$-axis points along geodetic zenith, the $y$-axis lies in the plane defined by the center of the Earth, the Sun, and the satellite (perpendicular to $z$), and the $x$-axis is defined by the right-hand rule. The radiator panel's two opposing faces are nominally oriented along the $\pm x$-axis in this frame, making the radiator normal perpendicular to both the zenith direction and the Sun--Earth--satellite plane. Because this frame is defined relative to the instantaneous Sun and satellite geometry, the radiator orientation evolves continuously as the satellite moves along its orbit, even though the radiator does not independently track the Sun in the way the solar array does. To account for mechanical flexing of the panel during operations, we apply a small wobble of $5^\circ$ to each face normal, tilting it within the plane orthogonal to the nominal axis. While the radiator is modeled as a purely diffuse (Lambertian) reflector rather than a specular one, its large surface area means the wobble can still produce a clearly visible effect on the resulting brightness.

A key assumption in the model concerns solar array orientation. Rather than assuming that the panel normal is perfectly aligned with the Sun direction, we offset the panel normal by $20^\circ$ from the Sun toward nadir. This is common practice, since solar panel efficiency decreases over the lifetime of a satellite. This effect can be compensated for by using a slightly larger array that is slightly off-pointed initially. The panel normal is therefore geometry-dependent and evolves with the Sun direction, but does not satisfy perfect Sun-tracking. For simplicity, the same attitude prescription is applied uniformly across all constellation shells, including both low-inclination and Sun-synchronous populations. The radiator orientation is therefore also geometry-dependent, but unlike the solar array, it is not actively Sun-tracking; rather, its orientation follows passively from the satellite-centered frame definition as the satellite orbits.

For each satellite position that is both within the field of view and sunlit, we compute the observer-frame reflected intensity at the Rubin site using the satellite altitude, satellite azimuth and elevation, solar azimuth and elevation, and the adopted surface BRDF models for the bus, solar array, and radiators. Direct solar illumination is included; earthshine is neglected. The resulting intensity \(I\) is then converted into an instantaneous apparent AB magnitude using
\[
m_{\rm AB}
=
-2.5 \log_{10}
\left(
\frac{I\, \lambda}
{c\, F_0}
\right),
\]
where \(\lambda = 532\,\mathrm{nm}\) is the adopted wavelength, \(c \approx 3\times10^8\,\mathrm{m\,s^{-1}}\) is the speed of light, and \(F_0=3631\,\mathrm{Jy}=3631 \times 10^{-26}\,\mathrm{Wm^{-2}Hz^{-1}}\) defines the AB magnitude zero point. The wavelength of \(532\,\mathrm{nm}\) is adopted because the laboratory-measured BRDF models used in this work were characterized at that wavelength. This yields a time-resolved apparent brightness estimate for every visible satellite crossing. Because the calculation is restricted to the sunlit in-field subset, the brightness model is applied only where it is physically relevant and observationally meaningful.

This framework allows us to capture two qualitatively different brightness regimes: a diffuse baseline associated with ordinary scattering from the bus, array, and radiator surfaces, and a brighter tail produced when the Sun--satellite--observer geometry approaches the specular condition for the bus or solar array. As a result, the model provides not only an estimate of how many LSST pointings are affected, but also how severe those impacts may be in terms of optical brightness.

\subsection{Stage 5: Streak Surface Brightness}\label{sec:streakSB}

The final stage of the pipeline converts the monochromatic AB mag estimates into per-band streak surface brightness measurements expressed in mag\,arcsec$^{-2}$. This conversion is necessary because Rubin's LSSTCam photometric system operates across six broad filters ($ugrizy$) and satellite streak profiles are very wide due to the defocus effect \citep{bektesevic18}. The impact of a satellite streak on science data depends not only on how bright the satellite is in total, but on how concentrated that brightness is per unit solid angle on the detector.

\subsubsection{Solar Color Correction}\label{solar_correction}
The \textsc{Lumos-Sat} framework computes a monochromatic AB magnitude at a fixed reference wavelength of 532\,nm. To translate this into the AB magnitude that LSSTCam would measure in a given filter band $b$, we apply a solar color correction:
\begin{equation}
m_b = m_{532} + \Delta_b^{\odot},
\end{equation}
where $\Delta_b^{\odot}$ is the difference in AB magnitude between the band-averaged solar flux and the solar flux at exactly 532\,nm:
\begin{equation}
\Delta_b^{\odot} = -2.5\log_{10}\!\left(\frac{\langle F_\nu \rangle_b}{F_\nu(532\,\mathrm{nm})}\right).
\end{equation}
Here $\langle F_\nu \rangle_b$ is the solar spectral flux density averaged over the filter bandpass, and $F_\nu(532\,\mathrm{nm})$ is the solar flux at the reference wavelength.
Both are evaluated using the solar spectral irradiance atlas distributed with \texttt{rubin\_sim} \citep{yoachim2022}, convolved with the Rubin total system throughput curves from 
\texttt{syseng\_throughputs} v1.9\footnote{\url{https://github.com/lsst-pst/syseng_throughputs}}, which include contributions from the atmosphere, mirrors, filter, and detector quantum efficiency. This correction assumes a gray (spectrally flat) satellite reflectance, so that the satellite spectrum mirrors the solar spectrum. Applying the method described above yields 
$\Delta_b^{\odot} = +1.43, +0.25, -0.21, -0.32, -0.36, -0.37$\,mag 
for the $u$, $g$, $r$, $i$, $z$, and $y$ bands, respectively 
(Table~\ref{tab:bandparams}). The positive value in $u$-band reflects the Sun's relatively low flux at short wavelengths compared to a spectrally flat AB source, 
while the $r$ through $y$ bands show modest negative corrections as the solar flux rises toward redder wavelengths.

\subsubsection{Photon Counting Formula}
We convert the corrected band magnitude $m_b$ into a detected electron count using a photon-counting model evaluated at the effective wavelength $\lambda_b$ and effective bandwidth $\Delta\lambda_b$ of each filter. The effective bandwidth is defined as
\begin{equation}
\Delta\lambda_b = \frac{\int T_b(\lambda)\,\mathrm{d}\lambda}{T_b(\lambda_b)},
\end{equation}
where $T_b(\lambda)$ is the total system throughput curve (atmosphere, mirrors, filter, and detector quantum efficiency) and $\lambda_b$ is the throughput-weighted effective wavelength. This definition ensures that the monochromatic approximation at $\lambda_b$ reproduces the correct integrated photon count.

Next, we compute the number of detected electrons accumulated during the satellite crossing in three steps. First, the band-corrected AB magnitude is converted to a spectral flux density,
\begin{equation}
    F_\nu(m_b) = 10^{-m_b/2.5} \times 3631 \times 10^{-26}
    \quad [\mathrm{W\,m^{-2}\,Hz^{-1}}].
\end{equation}
Second, $F_\nu$ is converted to an integrated in-band intensity. Since $F_\nu$ is defined per unit frequency, it must first be converted to $F_\lambda$ (flux per unit wavelength) using the standard Jacobian $|d\nu/d\lambda| = c/\lambda^2$, which follows from $\nu = c/\lambda$, where $c$ is the speed of light:
\begin{equation}
    F_\lambda = F_\nu \cdot \frac{c}{\lambda_b^2}
    \quad [\mathrm{W\,m^{-2}\,m^{-1}}].
\end{equation}
The total in-band intensity is then obtained by integrating $F_\lambda$ over each effective filter bandwidth $\Delta\lambda_b$:
\begin{equation}
    I_b = F_\lambda \cdot \Delta\lambda_b 
        = F_\nu \cdot \frac{c}{\lambda_b^2} \cdot \Delta\lambda_b
    \quad [\mathrm{W\,m^{-2}}].
\end{equation}
Third, the intensity is converted to a photon flux and multiplied by 
the collecting area, throughput, crossing time, and PSF fraction to 
give the total detected electron count,
\begin{equation}
    N_e = \frac{I_b}{hc/\lambda_b}
          \cdot A_{\rm eff}
          \cdot T_b(\lambda_b)
          \cdot t_{\rm cross}
          \cdot f_{\rm PSF},
\end{equation}
where $h$ is the Planck constant, $A_{\rm eff} = 33.1$\,m$^2$ is the Rubin mean effective aperture (area-weighted over the field of view)\footnote{\url{https://rubinobservatory.org/for-scientists/rubin-101/key-numbers}\label{fn:rubinkey}}, $T_b(\lambda_b)$ is the total system throughput at the effective wavelength $\lambda_b$ (atmosphere + mirrors + filter + detector QE), $t_{\rm cross}$ is the time the satellite spends crossing the field of view, measured directly from the propagated trajectory and $f_{\rm PSF} = 0.954$ is the fraction of flux captured within $\pm2\sigma$ of a Gaussian PSF. The throughput values $T_b(\lambda_b)$ and effective bandwidths $\Delta\lambda_b$ for all six LSSTCam filters are summarized in Table~\ref{tab:bandparams}.

\subsubsection{Streak Surface Brightness}
The mean number of electrons per pixel along the streak is
\begin{equation}
e_{\rm px} = \frac{N_e}{L_{\rm px}\,\times\,1.698\,\theta_{\rm eff}/s_{\rm px}},
\label{number_ele}
\end{equation}
where $L_{\rm px}$ is the streak length in pixels, $s_{\rm px} = 0.2$\,arcsec\,px$^{-1}$ is the Simonyi plate scale, $1.698$ is $4\sigma/\mathrm{FWHM}$ converts the effective PSF full-width at half-maximum $\theta_{\rm eff}$ into the $4\sigma$ cross-track footprint width, and $\theta_{\rm eff} = \sqrt{\theta_{\rm PSF}^2 + \theta_{\rm def}^2}$ combines the nominal Rubin PSF ($\theta_{\rm PSF} = 0.67''$) in quadrature with the defocus blur $\theta_{\rm def} = 206265\,D/s$, where $D = 8.36$\,m is the Rubin primary mirror diameter and $s$ is the satellite slant range in meters (straight-line line-of-sight distance from telescope to satellite). At the orbital altitudes considered here (686--1002\,km), the defocus blur ranges from approximately 1.7$''$ to 2.5$''$, substantially broadening the streak cross-section relative to a point source.

The streak surface brightness is then
\begin{equation}
\mu_b = \mathrm{ZP}_b - 2.5\log_{10}\!\left(\frac{e_{\rm px}/t_{\rm cross}}{s_{\rm px}^2}\right)\quad[\mathrm{mag\,arcsec}^{-2}],
\end{equation}
where $\mathrm{ZP}_b$ is the photometric zeropoint for band $b$ \citep{SMTN-002}, and $s_{\rm px}^2 = 0.04$\,arcsec$^2$ is the solid angle of sky seen by one Rubin pixel. We note the zeropoint is used here only for the final unit conversion from electrons\,s$^{-1}$\,arcsec$^{-2}$ to mag\,arcsec$^{-2}$; all photon counting is performed using the throughput-based formula above.

\subsubsection{Detectability}
A satellite streak is considered detectable if its band-corrected AB magnitude satisfies:
\begin{equation} \label{eq10}
m_b < m_{b,5\sigma},
\end{equation}
where $m_{b,5\sigma}$ is the single-visit $5\sigma$ point-source depth for band $b$ from \citet{pstn054}: $m_{5\sigma} = 23.8, 24.5, 24.0, 23.4, 22.7, 22.0$\,mag for $u$, $g$, $r$, $i$, $z$, and $y$ bands, respectively. We note that these are point-source limits, whereas satellite streaks are spatially extended \citep[as a function of orbital altitude, as demonstrated in, e.g.,][]{snyder25}, so this criterion is optimistic. We quantified this directly: across every streak in our simulated 365-day window, $91.0\%$ of the streaks passing this point-source criterion also clear the more physically appropriate surface-brightness threshold used throughout the rest of this work ($\mu_b < 24.5\,\mathrm{mag\,arcsec^{-2}}$, Section~\ref{sec:pixelLoss}), while the remaining $9.0\%$ ($801{,}053$ of $8{,}940{,}149$ streaks) are over-classified as ``detectable'' by the point-source comparison alone. This over-classification is real but modest for our simulated population, and it does not propagate into the pixel-loss results presented later in this paper: Equation~\ref{eq10} is a descriptive comparison only and plays no role in the pixel-loss processing pipeline itself, where $\mu_b$ -- not $m_b$ -- governs which streaks are treated as below the detection threshold (``faint'') in the saturation classification of Section~\ref{sec:pixelLoss} onward. Nonetheless, as a first-order indicator of whether a streak would manifest in Rubin images, this comparison offers a straightforward and observationally grounded threshold.

\begin{deluxetable}{lccccc}
\tablecaption{Photometric parameters used to compute band-specific streak surface brightnesses.}
\tablewidth{0pt}
\tablehead{
  \colhead{Band} &
  \colhead{$\lambda_b$ (nm)} &
  \colhead{$T_b(\lambda_b)$} &
  \colhead{$\Delta\lambda_b$ (nm)} &
  \colhead{$\mathrm{ZP}_b$ (mag)} &
  \colhead{$\Delta_b^{\odot}$ (mag)}
}
\startdata
$u$ & 366 & 0.140 & 55.9  & 26.52 & $+1.43$ \\
$g$ & 482 & 0.459 & 142.4 & 28.51 & $+0.25$ \\
$r$ & 622 & 0.536 & 140.5 & 28.36 & $-0.21$ \\
$i$ & 754 & 0.622 & 124.8 & 28.17 & $-0.32$ \\
$z$ & 869 & 0.613 & 102.1 & 27.78 & $-0.36$ \\
$y$ & 971 & 0.318 &  90.2 & 26.82 & $-0.37$ \\
\enddata
\tablecomments{Because our satellite brightness estimates are monochromatic, we use these parameters to compute streak surface brightnesses $\mu_b$ in mag\,arcsec$^{-2}$ for each Rubin Observatory band. The table lists each band's effective wavelength $\lambda_b$, total system throughput at that wavelength $T_b(\lambda_b)$, effective bandwidth $\Delta\lambda_b$, photometric zeropoint $\mathrm{ZP}_b$, and solar color correction $\Delta_b^{\odot}$ applied to translate our 532\,nm brightness model output into Rubin $ugrizy$ bands.\label{tab:bandparams}}
\end{deluxetable}

\subsection{Stage 6: Streak Pixel Loss}\label{sec:pixelLoss}

Having computed the streak surface brightness $\mu_b$ for each satellite crossing, we finally estimate the fraction of focal-plane pixels rendered unusable by that crossing. The calculation proceeds in three stages: classification of the streak by brightness, construction of a masked pixel footprint, and aggregation over all streaks within a pointing. While pixel loss is not the only metric that should be used to assess scientific impact \citep{lsst2025-report}, it is still an illustrative quantity.

\subsubsection{Streak Classification}\label{Streak Classification}

The electron count per pixel $e_{\mathrm{px}}$ computed in Stage~5 (Equation \ref{number_ele}) is used directly to classify each streak.
The peak electron count in the brightest pixel of the cross-track profile is
\begin{equation} \label{eq:electron peak}
    e_{\mathrm{peak}} = e_{\mathrm{px}} \cdot f_{\mathrm{peak}},
\end{equation}
where $f_{\mathrm{peak}}$ is the fraction of total streak flux concentrated in the peak pixel of the point-spread function, evaluated for the LSST median seeing of $0.67''$ at the Simonyi plate scale of $0.2''\,\mathrm{px}^{-1}$$^{(\ref{fn:rubinkey})}$. Each streak is then assigned to one of three categories based on $e_{\mathrm{peak}}$ and the LSSTCam full-well capacity $e_{\mathrm{FW}} = 130{,}000\,\mathrm{e^-}$ \citep{giraldo2023rtn}:

\begin{itemize}
    \item \textbf{Faint}: $e_{\mathrm{px}}$ is undefined because the streak falls below the point-source detection limit; no pixels are masked.
    \item \textbf{Unsaturated}: $e_{\mathrm{peak}} \leq e_{\mathrm{FW}}$; a cross-track mask of $W = 100$\,px (20 arcsec) is applied as a conservative estimate of the affected cross-track width. This is consistent with findings for fainter streaks in \citet{hasan22}.
    \item \textbf{Saturated}: $e_{\mathrm{peak}} > e_{\mathrm{FW}}$; pixel-well overflow produces bleed trails and charge-transfer artifacts, so a wider mask of $W = 500$\,px (100 arcsec) is applied to conservatively account for bleed extent and charge smearing. This is consistent with findings for brighter streaks in \citet{hasan22}.
\end{itemize}

We note a simplification in this classification: $f_{\rm peak}$ is a peak-to-average ratio calibrated for the delivered point-spread function core of a static point source -- that is, for the narrow, undiluted $\theta_{\rm PSF}$ profile, not for the wider, defocus-broadened $\theta_{\rm eff}$ profile (Equation~\ref{number_ele}) that satellite streaks at these orbital altitudes actually have. Because a broader profile is generically flatter than a narrower one carrying the same total flux, applying this point-source-calibrated factor to a density already diluted by the wider $\theta_{\rm eff}$ likely overestimates the true peak concentration, biasing Equation~11's saturation classification toward more streaks being classified as saturated than would occur in reality. Correctly resolving this would require either new instrumentation-team characterization of the peak-to-average ratio as a function of profile width, or a dedicated calibration effort beyond the scope of this work; we flag it here as a limitation of the current saturation classification.

\subsubsection{Streak Footprint and Pixel Loss} \label{Streak Footprint and Pixel Loss}
For each streak that is not fainter than the detection limit (``detectable streak''), the satellite trajectory is projected onto the focal plane using a gnomonic projection centered on the pointing, with the camera orientation accounted for via the \texttt{rotSkyPos} angle from the \texttt{opsim} survey database $^{\ref{opsim}}$. The physical camera frame is related to the sky-north projection by a rotation of $\varphi = 180\degr - \mathrm{RotSkyPos}$, applied as 
\begin{align}
    x_{\mathrm{cam}} &=  \cos\varphi\, x_{\mathrm{sky}}
                        + \sin\varphi\, y_{\mathrm{sky}}, \\
    y_{\mathrm{cam}} &= -\sin\varphi\, x_{\mathrm{sky}}
                        + \cos\varphi\, y_{\mathrm{sky}},
\end{align}
where $(x_{\mathrm{sky}}, y_{\mathrm{sky}})$ are the gnomonic projection coordinates in pixels ($+Y$\,=\,North, $+X$\,=\,East) and $(x_{\mathrm{cam}}, y_{\mathrm{cam}})$ are the final camera-frame coordinates. The $180\degr$ offset arises because the odd number of mirrors in the Rubin Telescope rotates the projected image by $180\degr$ relative to the sky \citep{SMTN-019}.

The masked footprint of each streak is modeled as a rectangle of length $L_{\mathrm{px}}$ (the streak length in pixels) and width $W$ (the mask width for its saturation class). The area of this rectangle in pixels is
\begin{equation}
    A_{\mathrm{streak}} = L_{\mathrm{px}} \times W.
\end{equation}

Because multiple satellites may cross the same pointing within a single 30\,s exposure, their footprints can overlap. We consider two approaches to combining overlapping footprints, which differ in whether they account for the additional electrons a second streak deposits in a region already covered by another streak's mask.
 
\paragraph{Fast (union-only) approach.} 
In this approach, each streak's saturation class -- and therefore its mask width $W$ -- is fixed by its own brightness in isolation (Section~\ref{Streak Classification}), with no accounting for the electrons contributed by any other streak sharing the same
pixels. The union of all streak mask rectangles within a pointing is computed using exact polygon geometry (\texttt{shapely}; \citealt{shapely2007}), which prevents double-counting of area shared by more than one streak but does not change any streak's assigned saturation class. The normal-streak pixel loss under this approach is
\begin{equation}
    A_{\rm normal}^{\rm fast} = \left| \bigcup_s M_s \right|,
    \label{eq:fast-union}
\end{equation}
where $M_s$ is streak $s$'s mask polygon at its own,
individually-determined width.
 
\paragraph{Slow (electron-summing) approach.}
Two individually unsaturated streaks crossing the same region can jointly deposit enough charge to saturate that shared area, an effect the fast approach does not capture. We account for this by summing electron counts in any region where two streaks' masks physically intersect, and re-evaluating the saturation class of \emph{only that overlapping sub-region} against the same full-well threshold $e_{FW}$ used in Section~\ref{Streak Classification}. Concretely, for every pair of detectable normal streaks $(i,j)$ whose mask polygons intersect, we compute the combined peak
electron count 
\begin{equation}
    e_{\rm peak}^{(i,j)} = e_{\rm peak}^{(i)} + e_{\rm peak}^{(j)},
    \label{eq:combined-electrons}
\end{equation} 
using the per-streak peak electron counts from Equation~\ref{eq:electron peak}, and reclassify the intersection region alone as saturated if $e_{\rm peak}^{(i,j)} > e_{FW}$, even when neither streak individually crosses that threshold. Only the geometric intersection of the two masks -- not the full length of either streak -- is widened to $W = 500$\,px; the remainder of each streak keeps the mask width set by its own individual classification, so a streak that overlaps another for only a short stretch of its trajectory is not conservatively widened along its entire length. As in the fast approach, the union of all (now segment-wise variable-width) mask polygons is computed with shapely, giving
\begin{equation}
    A_{\rm normal}^{\rm slow} = \left| \bigcup_s M_s' \right|,
    \label{eq:slow-union}
\end{equation}
where $M_s'$ is streak $s$'s mask polygon after any overlap-driven segment-wise widening. By construction $A_{\rm normal}^{\rm slow} \geq A_{\rm normal}^{\rm fast}$ for every pointing.

Unless otherwise noted, all pixel-loss results reported in this paper use the slow (electron-summing) approach. Although more computationally expensive than the fast approach, we adopt it as our primary result throughout because it more completely captures the additional saturation produced when multiple streaks overlap, giving a more accurate estimate of the true pixel loss. 

It is important to note that the pixel loss calculation completely neglects numerous faint streaks that fall below the detection threshold but which may still contribute considerable systematic error to background model estimation and source detection and measurement, which is beyond the scope of this work.

\subsubsection{Treatment of Extremely Bright Satellites}

A small number of satellites are sufficiently bright that the streak cannot be characterized by a surface-brightness model at all. We define an ``ultra-bright'' threshold at $m_{b} < 2$\,mag (AB), corresponding to satellites that would be visible to the naked eye. For such streaks, the standard photon-counting model breaks down and the precise pixel loss depends strongly on instrument-specific effects that are not modeled here.

We therefore adopt a conservative upper-bound approach: every CCD that the satellite trajectory intersects is declared entirely lost. This is consistent with \citet{polin26}, and has been empirically verified with a handful of overwhelmingly bright streaks seen in early Rubin Operations images (see Appendix \ref{sec:superbright}).

The Rubin focal plane contains 189 science CCDs, each with $4096 \times 4096$ pixels. The CCD boundaries are computed in the camera-frame pixel coordinate system described above, and a satellite trajectory segment is intersected with each CCD bounding box by dense interpolation (50 steps per segment). 

All CCDs hit by an ultra-bright streak are declared entirely lost. The total pixel loss for a pointing combines this dead-CCD area with the normal-streak mask area (computed via either the fast or the slow approach of Section~\ref{Streak Footprint and Pixel Loss}) as a single unified region, rather than as independently-computed areas added together: the union of every dead CCD's full pixel footprint together with the normal-streak mask union, intersected with the true science-pixel footprint of the focal plane $\mathcal{F}$ (excluding the inter-CCD, inter-raft, and corner wavefront-sensor gaps that carry no science pixels), gives the total lost area for the pointing:
\begin{equation}
    A_{\rm total} = \left( \bigcup_{c \,\in\, \rm dead} C_c \;\cup\; \bigcup_s M_s^{(\ast)} \right) \cap\, \mathcal{F},
    \label{eq:combined-union}
\end{equation}
where $C_c$ is the full pixel footprint of dead CCD $c$, and $M_s^{(\ast)}$ denotes streak $s$'s mask polygon under whichever approach (fast, $M_s$, or slow, $M_s'$) is being reported. The pixel loss fraction for the pointing is then
\begin{equation}
    f_{\rm loss} = \frac{A_{\rm total}}{N_{\rm total}},
    \label{eq:floss-new}
\end{equation}
which by construction cannot exceed unity, since $A_{\rm total}$ is itself a subset of the focal-plane footprint used to define $N_{\rm total}$.\footnote{We note $N_{\rm total}$ as defined geometrically from the 189-CCD footprint does not equal $189 \times 4096^2 = 3{,}170{,}893{,}824$ exactly, owing to a small difference between the nominal per-CCD pixel count and the CCD's physical footprint as projected onto the focal plane; we [reconcile / note] this discrepancy.}
 
\section{Results}\label{sec:results}

Our aim was to answer four questions: (1) how often are LSST pointings intersected by satellites, (2) where do visible satellites appear on the Rubin sky, (3) how bright are those visible satellites, and (4) how does that brightness vary with season and time of night. We address each of these questions in turn.

\subsection{Fraction of Pointings Affected by Satellite Streaks}

\begin{figure*}[!ht]
\centering
\includegraphics[width=\textwidth]{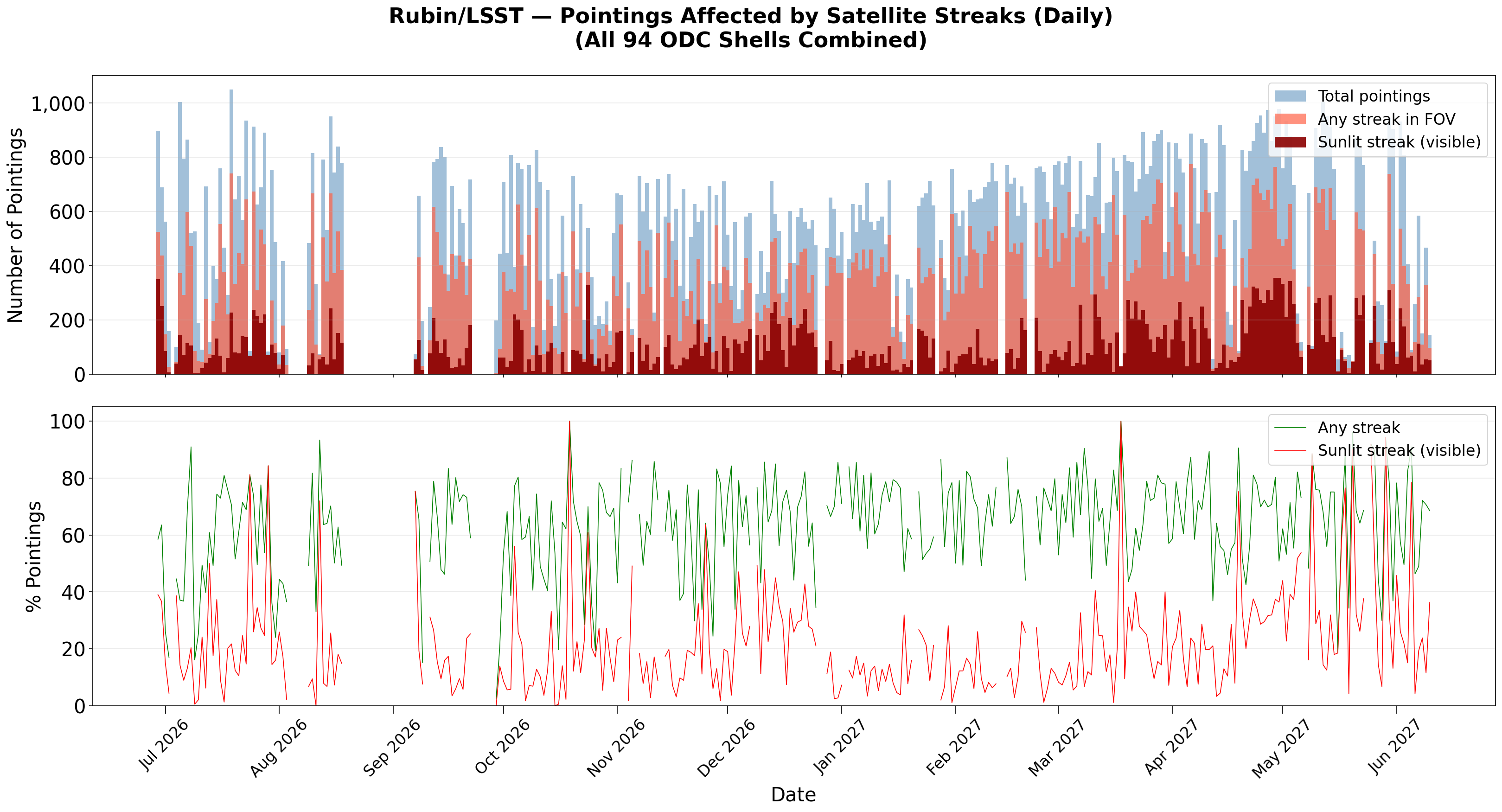}
\caption{
Nightly fraction of Rubin Observatory pointings affected by satellite streaks over one year of simulated LSST operations. The top panel shows the number of pointings per night broken down into total pointings, those intersected by any satellite streak, and those intersected by a sunlit (optically visible) satellite streak. The lower panel shows the same quantities expressed as a percentage of nightly pointings. Gaps in the time series correspond to periods with no scheduled observations in the adopted simulation, including planned downtime and weather-related losses.
}
\label{fig:streak_fraction}
\end{figure*}

We first quantify the fraction of exposures affected by satellite streaks on a per-Julian-day basis during one year of scheduled LSST observations. Figure~\ref{fig:streak_fraction} shows the nightly number of pointings affected by satellites as well as percentages of pointings crossed by sunlight satellites and by those shadowed by the Earth.

Our results suggest that, on average, more than half of LSST pointings will contain at least one satellite, corresponding to a mean of $\sim$180,000 geometric satellite crossings and $\sim$60,000 illuminated streaks per observing night. This result is not surprising given the enormous number of satellites and broad distribution of shell inclinations and altitudes shown in Table~\ref{tab:table1}. However, the geometric crossing rate alone overstates the impact on LSST observations. Only a subset of these crossing satellites are sunlit and, therefore, capable of producing visible optical streaks (the impact of non-sunlit satellites occulting astrophysical sources is not considered here), as shown in the lower panel of Figure \ref{fig:streak_fraction}.

The temporal variability of satellite crossings reflects the strong seasonal dependence of the Sun--satellite--observer geometry. Periods with a higher fraction of visible streaks correspond to times when the Sun remains favorably positioned to illuminate satellites while Rubin is operating under dark or near-dark observing conditions. Conversely, during deeper night, many satellites that still cross the field geometrically are located within Earth's shadow and do not produce visible trails in optical wavelengths. The modulation is thus driven primarily by changes in the Sun--satellite--observer geometry rather than by changes in orbital occupancy.

Even after accounting for illumination constraints, the visible impact remains substantial throughout the year. During some observing periods, more than \(80\%\) of LSST pointings contain at least one illuminated satellite streak, while the average level of contamination hovers around 20\%. This demonstrates that mitigation strategies, such as lowering orbits, can reduce, but do not eliminate, the observational impact of a million-satellite-scale orbital data center constellation. 

The strong seasonal modulation further implies that visible satellites are not uniformly distributed across the sky. Instead, the locations of illuminated satellites evolve systematically with the Sun--satellite--observer geometry, producing preferred regions of sky contamination that vary throughout the year. In the following subsection we examine the sky distribution of these sunlit satellites in the Rubin observer frame.

\subsection{Sky Distribution of Sunlit Satellites}

\begin{figure}[!ht]
\centering
\includegraphics[width=0.79\columnwidth]{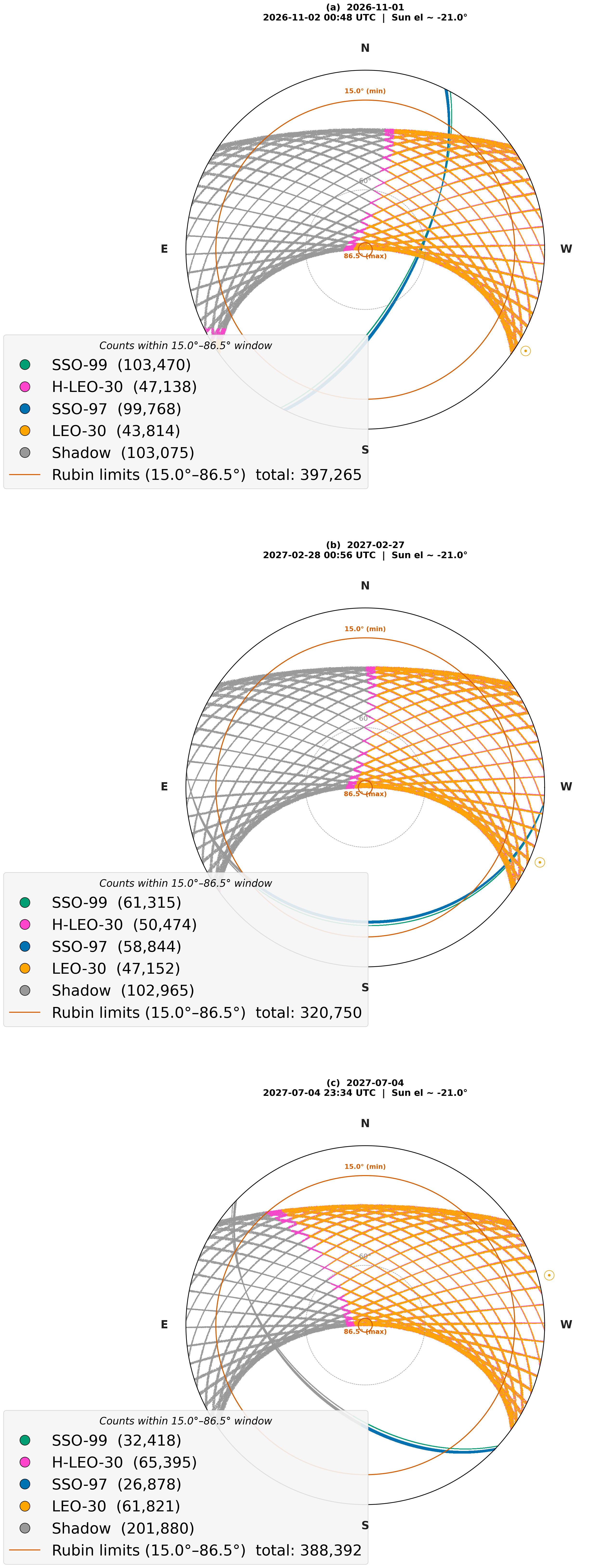}
\caption{
Sky distribution of satellites in the Rubin observer frame for representative dates at comparable solar elevation angles (\(\sim-21^\circ\)). Each point represents a satellite position in azimuth--elevation coordinates, with colored points corresponding to illuminated satellites and gray points indicating satellites in Earth's shadow. Satellite populations are grouped by orbital regime(see Table~\ref{tab:table1}). The orange circle marks the nominal Rubin observing elevation limits (\(15^\circ\)--\(86.5^\circ\)) adopted from LSST pointing constraints (\url{https://rubinobservatory.org/for-scientists/rubin-101/telescopes}).
}
\label{fig:sky_density}
\end{figure}

Figure~\ref{fig:sky_density} shows the sky distribution of satellites seen from Rubin Observatory for representative dates chosen to have similar solar elevation angles, approximately $-21^\circ$, i.e., near the beginning or end of the astronomical night. Each point corresponds to a satellite position, with illuminated satellites shown in color and satellites in Earth's shadow shown in gray.

Although satellites are present across a large fraction of the observer sky, only a localized subset are illuminated at any one time. In other words, the visible population occupies a restricted region of the sky defined by the simultaneous requirements that the satellite be above the local horizon, remain illuminated by the Sun, and be visible from Rubin Observatory at the relevant observing geometry. Most of the remaining sky is occupied either by shadowed satellites or by no satellites at all.

This anisotropic distribution means that the burden on Rubin is not spatially uniform. Some parts of the sky are systematically more vulnerable to visible streaks than others, even when the total number of satellites above the horizon is large. The concentration of sunlit satellites into a limited region of the sky also suggests that the contamination risk is strongly directional and therefore tied to both the telescope pointing strategy and the seasonal placement of the illuminated orbital arcs.

The location and morphology of the region containing illuminated satellites evolve with the season. Even at comparable solar elevation, the illuminated band shifts in azimuth and elevation because the underlying Sun--satellite--observer geometry changes over the year. 
The fewest SSO satellites are visible in the Southern Hemisphere winter (July), and the affected region of the sky does not overlap at all with the LEO satellite regimes.

Naturally, the solar elevation angle plays a major role also. The farther the Sun is below the local horizon, the smaller the number of illuminated low-inclination LEO satellites as those shells enter Earth's shadow. In contrast, the Sun-synchronous populations remain preferentially illuminated under many observing geometries due to their orbital alignment with the terminator. Consequently, the relative contribution of the SSO regimes becomes increasingly important during deeper night conditions, even as the total illuminated satellite population decreases.

\subsection{Satellite Brightness Distribution}

We next examine the brightness distribution of the visible satellite population. Satellites appear as bright streaks in telescope images due to their rapid orbital motion relative to the astrophysical background targets \citep{boley25}. In a first step we use the brightness model framework in \citet{fankhauser23} to estimate stationary equivalent flux 
(subsequently converted into AB magnitude, and later surface brightness) for each sunlit satellite that passes through the Simonyi field of view over the one year period shown in Figure~\ref{fig:streak_fraction}.

The brightness model computes reflected intensity at a single reference wavelength of 532\,nm, matching the wavelength at which the BRDF models in \citet{fankhauser23} are defined. To translate this monochromatic AB magnitude into the filter-integrated AB magnitude measured by Rubin's LSSTCam in a given filter, we apply the solar color correction described in Section~\ref{solar_correction}, which assumes gray (spectrally flat) satellite reflectance.

\begin{figure*}[!ht]
\centering
\includegraphics[width=\textwidth]{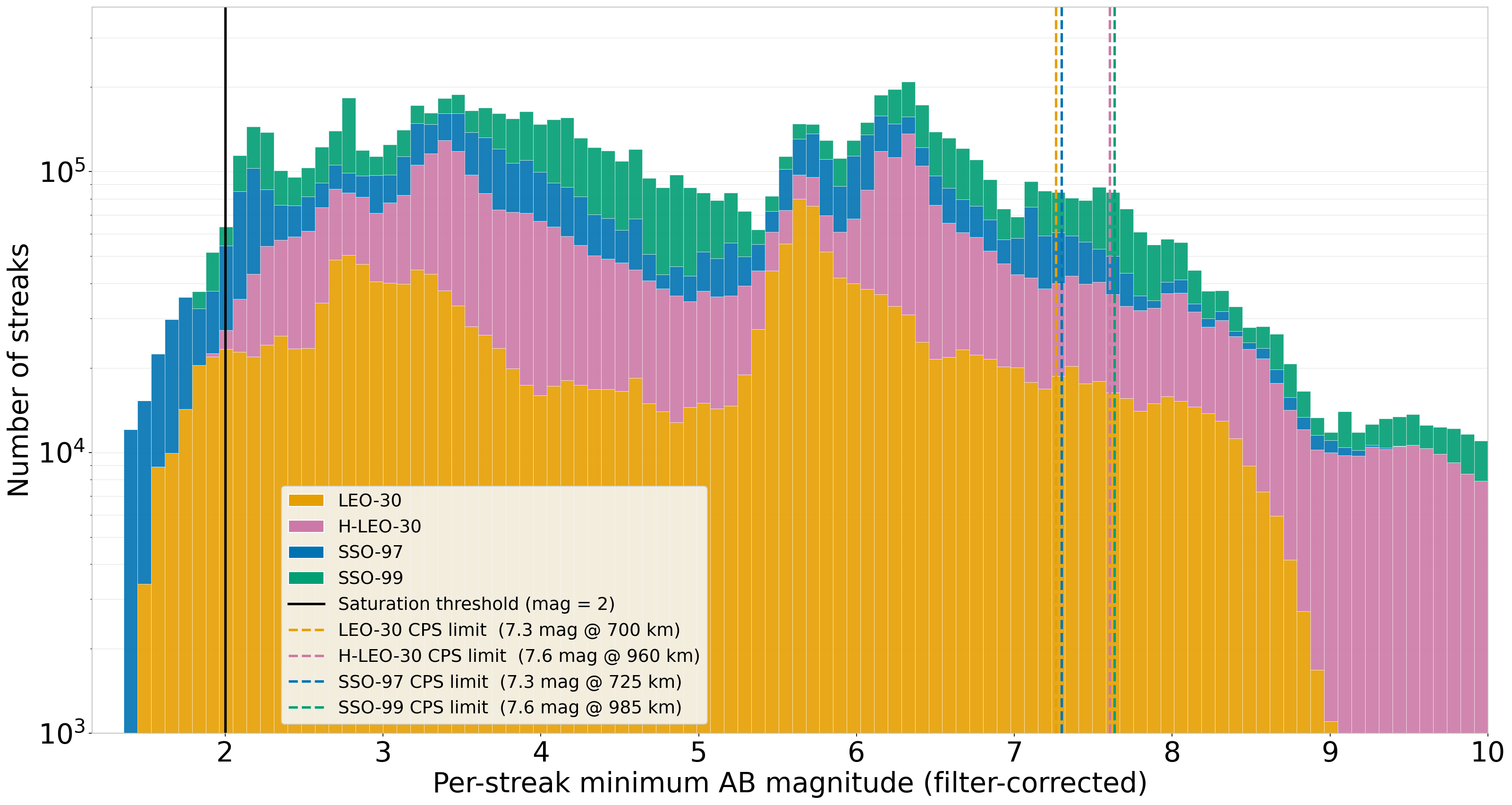}
\caption{
Distribution of per-streak brightest AB magnitude for all sunlit ODC constellation satellites within the Rubin Observatory Simonyi Survey Telescope field of view, aggregated across all simulated nights after
applying solar phase angle corrections to the AB magnitudes, including the contribution of the satellite radiator surface (Section~\ref{sec:brightness}). Bars are subdivided by orbital shell (colors follow Table 1); note the logarithmic y-axis scale. The peak at mag \(\sim 6\) is associated with typical diffuse and moderately directional reflection geometries, while the brighter peak at mag \(\sim 3\)--4 is consistent with near-specular reflection conditions involving the large solar arrays, enhanced by the diffuse reflective contribution of the radiator across a wide range of viewing geometries. The solid black vertical line marks the detector saturation threshold (mag = 2). Dashed vertical lines show the IAU CPS recommended brightness limit of mag = 7 (referenced to 550 km altitude), scaled to each shell's representative altitude \citep{eggl2025mitigating}.
}
\label{fig:brightness_hist}
\end{figure*}

Figure~\ref{fig:brightness_hist} shows the distribution of modeled satellite brightness expressed in AB magnitude. The distribution is multi--modal, 
indicating the complex scattering behavior of the ODC constellation.

Across the one-year simulation, we identify approximately \(9\times10^6\) sunlit satellite streaks brighter than or equal to AB magnitude 12.5 that intersect the Simonyi field of view. Approximately 83.9\% of modeled streaks are brighter than AB magnitude 7, which corresponds to the IAU-recommended limit \citep{boley25}, while more than 229,000 streaks reach magnitudes brighter than 2 (around 2.5\% of sunlit streaks).

The bright end of magnitudes of roughly 2--4 is due to near-specular reflection geometries involving the large ODC solar arrays. Although comparatively infrequent, such events dominate the brightest portion of the distribution and are important as they can cause detector saturation, nonlinear image artifacts, and substantial loss of Rubin data products.

The broader and more heavily populated peaks near magnitudes of 6 represent the characteristic brightness regime of the ODC constellation under more typical observing geometries. In this regime, the observed intensity is dominated by a combination of diffuse and moderately directional scattering from the satellite surfaces. Some additional structure near this dominant peak may arise from weaker specular reflections associated with the satellite chassis, whose reflective area is substantially smaller than that of the solar arrays and therefore produces fainter enhancements than the brightest glint events. Repeated Sun--satellite--observer viewing geometries may also contribute to the substructure in the predicted magnitude distribution, since the observing configurations accessible from Rubin are not sampled uniformly over time.

Including the radiator surface in the brightness model increases the apparent brightness of satellites across the population, with the magnitude of the increase depending on the instantaneous observer--satellite--Sun geometry. Because the radiator is modeled as a large-area, high-albedo ($0.9$) diffuse reflector, it contributes reflected flux toward the observer over a wide range of viewing geometries, rather than only near the narrow specular condition that produces the brightest tail from the bus and solar array. As a result, the radiator's contribution is visible throughout the distribution rather than being confined to the specular tail alone.

In summary, the ODC constellation likely produces both a persistent moderate-brightness population and a smaller but significantly brighter population, each of which may affect LSST observations in distinct ways.

\subsection{Seasonal and Diurnal Brightness Variations}\label{sec:seasonal}

\begin{figure*}[!ht]
\centering
\includegraphics[width=6in]{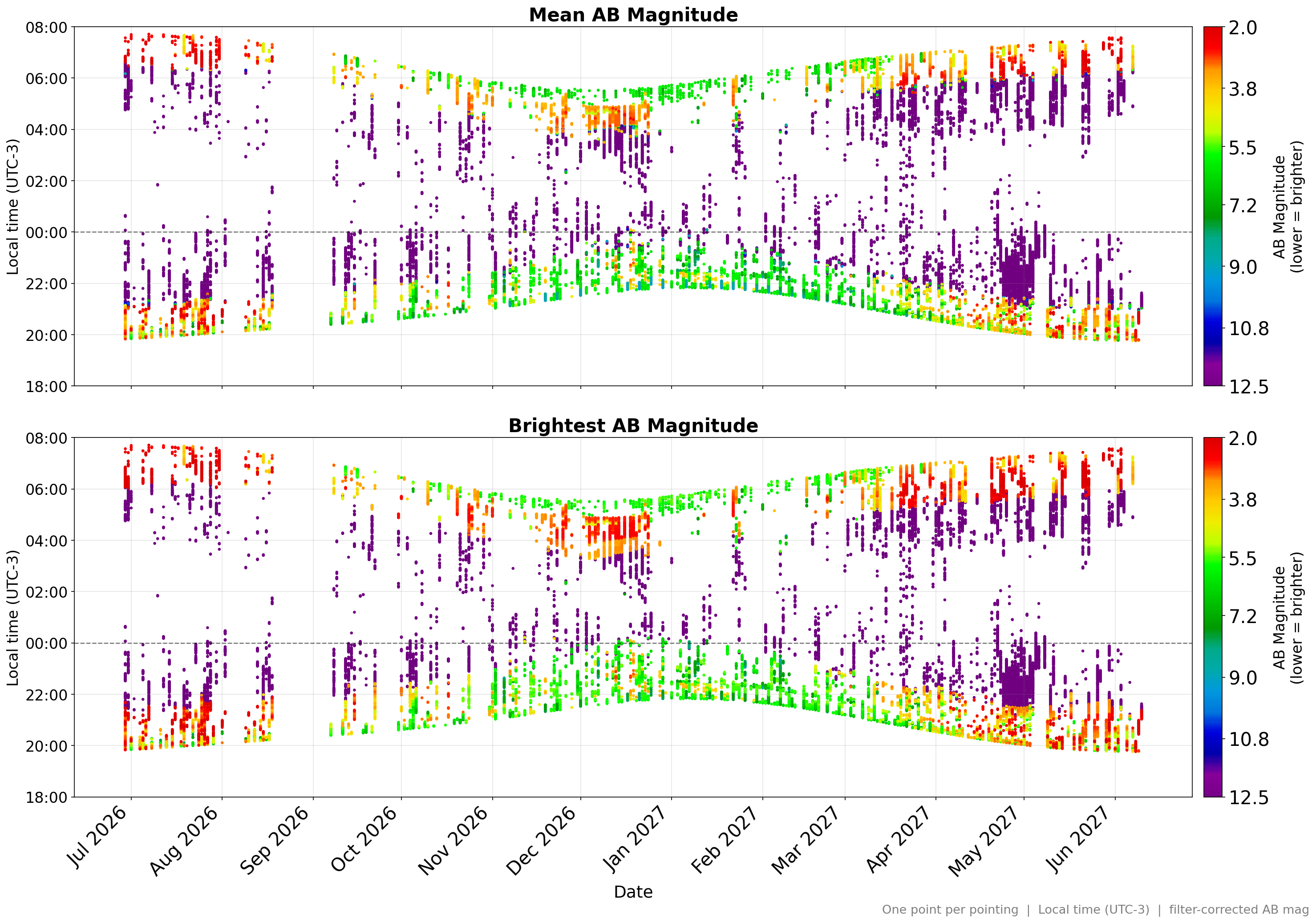}
\caption{
Temporal variation of modeled satellite brightness as a function of date 
and local time at Rubin Observatory after applying solar phase angle corrections to the AB magnitudes. The top panel shows the mean AB magnitude of visible satellites in each pointing; the bottom panel shows the brightest (minimum) AB magnitude per pointing, indicating the most severe contamination event in that interval. Red points correspond to brighter satellites (lower AB magnitude) and blue to fainter ones. White regions indicate times with no sunlit satellite crossings.
}
\label{fig:brightness_heatmap}
\end{figure*}

Figure~\ref{fig:brightness_heatmap} shows how the modeled satellite brightness varies as a function of date and time of night. The mean AB magnitude within each pointing is color-coded. The figure shows that satellite brightness varies with the season and observing time.

The brightest field-of-view crossing events cluster near twilight. This behavior follows directly from illumination geometry. Near twilight, many satellites remain sunlit even though the ground-based observer is already in darkness, creating the conditions for visible streaks. Under those circumstances, both the number of illuminated satellites and the probability of bright specular reflection are elevated.

As the night progresses and the Sun moves farther below the horizon, the population of illuminated satellites becomes both less numerous and typically fainter. The mean brightness, therefore, shifts toward fainter values during deep night, with the remaining visible population dominated by weaker, more diffuse reflection. However, this seasonal evolution is not constant throughout the year. The exact timing and intensity of the brightening bands drift with date, reflecting the annual cycle in solar declination and the corresponding change in Sun geometry at the Rubin site. This is also a strong function of latitude; Rubin is at a latitude of about $30^{\circ}$ S, and telescopes further from the equator will see different sky and brightness distributions of the same satellite population \citep{lawler22}.

White regions in Figure~\ref{fig:brightness_heatmap} indicate the absence of sunlit satellite crossings in a given time, which can result from three distinct causes: no satellite passing through the simulated field of view, all satellites present being in Earth's shadow, or no observation being scheduled at that time.

\subsection{Streak Surface Brightness}\label{sec:results_surface_brightness}

\begin{figure*}[!ht]
\centering
\includegraphics[width=6in]{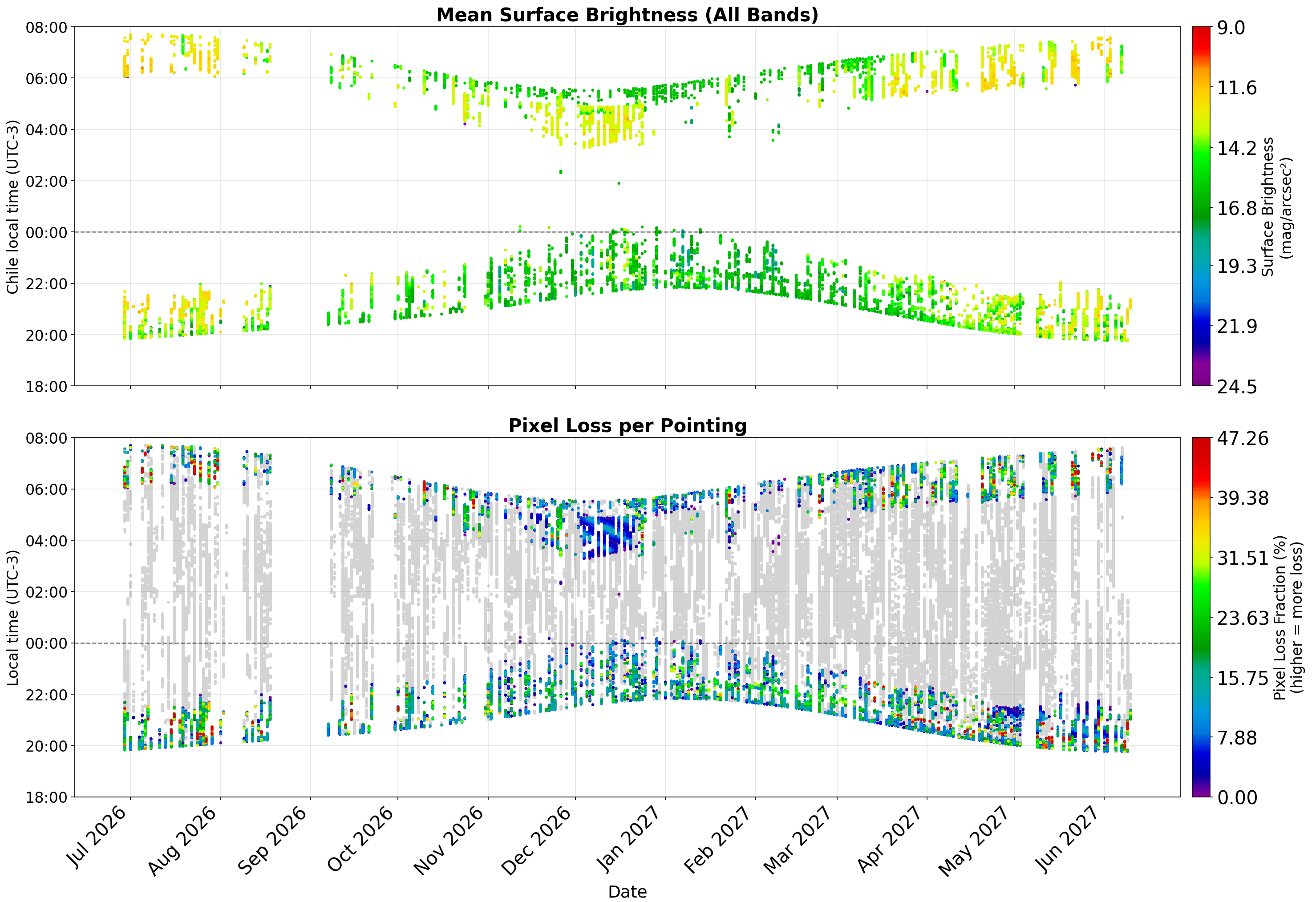}
\caption{
Top: Temporal variation of mean satellite streak surface brightness as a function of date and Chile local time (UTC-3), averaged over a pointing. The color scale runs from $\sim$9\,mag\,arcsec$^{-2}$ to 24.5\,mag\,arcsec$^{-2}$ ; white regions indicate times with no sunlit satellite crossings. The dashed horizontal line marks local midnight. Surface brightness is computed per LSSTCam band using the photon-counting formula described in Section~\ref{sec:streakSB}, accounting for per-band throughput, effective bandwidth, solar color corrections, satellite crossing
time, and defocus blur. Bottom: pixel loss fraction per pointing over the same time baseline, with each point colored by the fraction of pixels lost to satellite-streak masking; grey points indicate pointings with zero pixel loss
}
\label{fig:sb_heatmap}
\end{figure*}
Figure~\ref{fig:sb_heatmap} shows the modeled streak surface brightness averaged over a pointing as a function of date and time of night (top panel) and the resulting daily pixel loss for twilight pointings (bottom panel). Mean surface brightness values range from approximately 10 to 24.5\,mag\,arcsec$^{-2}$, while the brightest individual streaks reach as high as $\sim$9.94\,mag\,arcsec$^{-2}$, concentrated near twilight. The per-band surface brightness distributions are provided in Appendix~\ref{app:sb_bands} (Figure ~\ref{fig:sb_bands}).

The temporal pattern closely mirrors that of the AB magnitude heatmap (Figure~\ref{fig:brightness_heatmap}), with the brightest surface brightness events concentrated near twilight. However, the two quantities are not equivalent: surface brightness measures flux per unit solid angle on the detector and therefore depends not only on the satellite's total reflected flux but also on how that flux is distributed along the streak. A satellite with a faint integrated magnitude but a short, fast-moving streak can produce a high surface brightness because its photons are concentrated into fewer pixels. This explains why some pointings near twilight show elevated surface brightness even when the integrated AB magnitude is relatively faint.

A streak is considered detectable if its surface brightness exceeds $24.5\,\mathrm{mag\,arcsec}^{-2}$, which accounts for the spatial dilution of flux across the streak footprint through the defocus blur and crossing time. This criterion is therefore more physically appropriate for extended sources than a simple point-source depth comparison. \citet{tyson24} demonstrates the importance of accounting for this dilution effect for smaller space debris objects.

\subsection{Pixel Loss}\label{sec:results_pixel_loss}
The practical impact on science data is captured more directly by the per-pointing pixel loss fraction shown in the bottom panel of Figure~\ref{fig:sb_heatmap}, computed using the slow (electron-summing) approach of Section~\ref{Streak Footprint and Pixel Loss} and including the contribution of the satellite radiator surface described in Section~\ref{sec:brightness}. Each point represents a single pointing, colored by its pixel loss fraction; grey points indicate pointings with zero loss. The colorbar is scaled to the 99th-percentile pixel loss fraction across all pointings, $47.26\%$, meaning roughly 1\% of pointings exceed this level of contamination. The maximum pixel loss in a pointing reaches $96.30\%$ (Figure~\ref{fig:pixelLossDistribution} and~\ref{fig:pixelLossTwilight})

It is worth emphasizing that these pointing-level loss fractions reflect the aggregate effect of many simultaneous streaks, not a single crossing. A na\"ive estimate built from one streak alone even accounting correctly for the plate scale and mask width established above -- underestimates the true per-pointing contamination by orders of magnitude, because affected pointings are rarely crossed by only one satellite. Restricting to twilight pointings, where streak activity is most concentrated (Section~3.2), the \emph{median} pointing contains approximately 100 simultaneous visible streaks, with the distribution's mean (294.9) pulled substantially higher by a heavy tail extending to pointings with several thousand simultaneous streaks. Combined with the $W=100\,\mathrm{px}$ ($20''$) and $W=500\,\mathrm{px}$ ($100''$) mask widths from Section~2.7.1 -- both considerably wider than a single streak's seeing-limited profile alone -- this streak multiplicity, not any single mechanism in isolation, is the dominant driver of the nightly mean pixel-loss levels reported in this section.

Pixel loss is consistently higher during twilight overall, driven both by the elevated streak surface brightness expected when satellites are sunlit and the sky background is still relatively dark, and simply by the far greater number of simultaneous streaks in a typical twilight pointing. Within the twilight window, loss is not uniformly distributed: the evening twilight band shows visibly more severe contamination than the pre-dawn band. Superimposed on this general twilight enhancement are sudden, isolated spikes to high pixel loss (the scattered red and orange points) that stand well above the surrounding pointings at the same time of night. These spikes are most plausibly explained by the pointing direction itself, rather than by an overall increase in satellite brightness or number: at those specific moments, the telescope is likely looking toward a region of sky where the constellation geometry causes satellites to be more spread out across the field of view, increasing the number of distinct streaks — and therefore the total pixel loss — within that single exposure. This pattern illustrates that if the pointing strategy does not account for the instantaneous distribution of satellites on the sky, pixel loss in an individual pointing can be substantially worse than the typical or median case. A more detailed analysis of the implications for science operations is discussed in Section~\ref{sec:harm}.

\subsection{Impact on Sky Brightness}\label{sec:sky_brightness}

The impact of ODCs on the overall sky brightness has been argued to have an equally if not greater detrimental impact on astronomy \citep{hainaut26}. We therefore investigate the effect of the same ODC constellation used to model pixel loss on the observed change in sky brightness in all LSST filters. As expected, we find the sky brightness increase is substantial, especially during New Moon. The increase in $u,g$ and $r$ filters can reach several magnitudes and more than 1000\% of the natural sky background during twilight hours and only slowly subsides toward the middle of the night. The $y$ band observations are less extremely affected, although the sky background still increases by 100\% during twilight. The presence of ODCs would even increase sky background during Full Moon nights by more than 100\% in $r,i,z$ and $y$ bands.
Details of our all-sky brightness modeling are described in Section \ref{sec:appendix-sky-brightness}.

\section{Mitigation Strategies}\label{sec:mitigation}

\citet{lsst2025-report} suggested several mitigation options that would reduce impact or constellations on astronomy. Among those were the deployment of fewer satellites, sun-synchronous orbit satellites only, and/or lowering the orbital altitude of satellites. In the following we discuss the potential benefit of such mitigation options. We do not consider the tradeoff between larger number of smaller satellites vs. fewer number of larger satellites because the simulated ODC constellation is already fundamentally a large number of large satellites. Neither do we consider operational avoidance, or ``satellite dodging,'' as in \citet{hu22}, because this strategy is only viable for at most a few thousand bright satellites.

\subsection{Fewer ODC Satellites}\label{sec:fewer}

In Section~\ref{sec:results}, we assume the full $\sim$1 million satellite constellation would be launched and deployed. While the impact from this scenario is unsurprisingly severe, we also wish to explore the regime where a fraction of the full ODC satellite constellation is operational. This would be a realistic scenario either if the total ODC size were scaled down or during the portion of Rubin operations when the ODC is being launched and deployed in phases.

The most important result from this scaling argument is that the impact does not disappear even when the constellation population is reduced substantially. Figure~\ref{fig:streak_fraction} demonstrates that the baseline geometric crossing rate is already extremely large for the simulated ODC architecture. Reducing the number of satellites would therefore primarily decrease the probability of simultaneous streaks and reduce the frequency of the brightest events, rather than eliminate contamination altogether.

Even at only 10\% deployment, the constellation would still contain approximately $10^5$ satellites in LEO, exceeding the scale of current-generation constellations by a substantial margin. Under such conditions, visible satellite streaks would remain common during twilight observing windows. Sky-distribution maps for this reduced population are shown in Appendix~\ref{app:tenth} (Figure~\ref{fig:sky_tenth}), confirming that the qualitative structure of the illuminated satellite population, including the relative contributions of each orbital regime is preserved at 1/10th scale.

Importantly, the impact does not scale purely with satellite number. Figure~\ref{fig:brightness_hist} shows that a significant fraction of modeled streaks are brighter than the IAU CPS recommended limit of 7th magnitude \citep{boley25}, with a bright tail extending toward detector saturation. Because these bright events are driven by specular reflection from large solar arrays, reducing the total number of satellites alone does not fully mitigate the most severe contamination events. This is further illustrated by the pixel loss analysis in Appendix~\ref{app:tenth} (Figure~\ref{fig:pixel_loss_tenth}): even at 10\% deployment, the mean twilight pixel loss is $\sim2.33\%$ per night, with individual pointings reaching up to $87.25\%$ loss. A 10-fold reduction in constellation size does not produce a 10-fold reduction in pixel loss, because a single bright streak is sufficient to dominate an entire pointing regardless of how many satellites were removed.

The implications for Rubin science therefore remain significant even in reduced-population scenarios. A partially deployed ODC constellation would still produce a persistent background of visible satellite streaks throughout the year, particularly during twilight conditions and in observing geometries favorable for Sun-synchronous illumination.

\subsection{Sun-Synchronous ODC Satellites Only}\label{sec:sso_only}

Our results from Section~\ref{sec:results} suggest that the ODC impact on Rubin is time-dependent: first through the probability that satellites are visible at all, and second through the magnitude distribution of those visible satellites. Bright streaks are therefore not distributed uniformly over the survey, but are preferentially concentrated into specific observing windows tied to twilight and season.

This pattern is already familiar for present-day LEO satellite constellations and other artificial objects in LEO (e.g., rocket bodies and debris). Where the simulated ODC differs strikingly between ``more of the same'' and a fundamentally new problem is its inclusion of two SSO components.

Figures~\ref{fig:sky_density} and~\ref{fig:brightness_heatmap} show that the Sun-synchronous shells remain preferentially illuminated under many twilight and near-twilight observing geometries. Unlike lower-inclination shells, which spend a larger fraction of nighttime inside Earth's shadow, the SSO shells maintain geometries that keep satellites illuminated even when the observer is in darkness. As a result, the SSO population contributes disproportionately to the visible contamination.

This behavior is particularly concerning because twilight observing periods are scientifically important for Rubin Observatory, especially for Solar System science applications such as near-Earth object discovery and characterization. During these observing windows, Figure~\ref{fig:brightness_heatmap} shows that both the frequency of visible satellites and the probability of bright reflection events are elevated.

The sky-distribution maps in Figure~\ref{fig:sky_density} further demonstrate that the illuminated SSO population occupies distinct regions of the observer sky that evolve seasonally with Sun geometry. In some seasonal configurations, the SSO contamination regions remain spatially separated from the low-inclination shells, while in others they contribute substantially to the visible sky density.

Consequently, even a scenario in which only the SSO component of the ODC constellation were deployed would still represent a severe challenge for Rubin operations. The orbital geometry of Sun-synchronous satellites makes them unusually efficient at producing visible optical contamination during the very observing periods most important for wide-field astronomical surveys.

\subsection{Lower Altitude ODC Satellites}\label{sec:lower_altitude}

Lowering satellite altitude is frequently proposed as a mitigation strategy because lower-altitude satellites spend more time inside Earth's shadow and are therefore visible for a smaller fraction of the night. In principle, reducing the orbital altitude of the ODC shells by several hundred kilometers would decrease the number of illuminated satellites visible from Rubin Observatory, particularly during deep night conditions.

This effect can already be inferred from Figures~\ref{fig:sky_density}. The low-altitude LEO shells exhibit a stronger reduction in visible population as the Sun moves farther below the horizon, whereas the higher-altitude and Sun-synchronous populations remain illuminated under a broader range of observing geometries. Lowering the constellation altitude would therefore likely reduce the persistence of visible contamination during the middle of the night.

However, lowering altitude does not fundamentally remove the observational challenge posed by a million-satellite-scale constellation. Figure~\ref{fig:streak_fraction} shows that the dominant issue is not only satellite visibility, but the extremely large geometric crossing rate associated with such a dense orbital population. Even if the illuminated fraction decreases, the underlying number of satellites crossing Rubin fields remains extremely high.

In addition, lower-altitude satellites move with larger apparent angular velocity across the detector. This can produce rapidly moving bright streaks that still contaminate images even if the total visible population is somewhat reduced. Furthermore, Figure~\ref{fig:brightness_hist} demonstrates that the brightest contamination events are driven by large reflective solar arrays and favorable specular geometries. Those reflection mechanisms would still occur at lower altitude, particularly during twilight conditions.

Operationally, lower altitudes also imply shorter orbital lifetimes due to atmospheric drag, requiring more frequent station-keeping maneuvers and replacement launches. For a constellation approaching one million satellites, this would substantially increase launch cadence and orbital traffic density while also raising concerns regarding orbital debris generation and disposal reliability.

Thus, while lower orbital altitude may reduce some aspects of nighttime visibility, it does not eliminate the fundamental scaling problem associated with extremely large populations of bright satellites in LEO.

\subsection{Additional Mitigation Strategies}

\textit{Operator-side specular-reflection avoidance.} Satellite operators can, in principle, actively orient solar panels and other reflective surfaces to avoid directing specular glints toward Cerro Pach\'on during Rubin's observing windows, analogous to existing coordination frameworks between satellite operators and the astronomical community for brightness mitigation. Because the highest-impact events in our simulation are driven disproportionately by the brightest glints rather than by diffuse-only reflection, operator-side glint avoidance targeted specifically at Rubin's site could suppress a meaningful fraction of the saturating-streak and ultra-bright-streak population without requiring any change to constellation size or altitude. This mitigation should nonetheless be understood as partial: it addresses only the brightest, specular-glint-driven spikes in the population, not the constellation's baseline brightness. As noted above, the simulated ODC constellation is fundamentally a large number of physically large satellites, and their diffusely reflected and thermally radiated flux -- independent of any specular glint -- scales with this physical size. Even with specular glints fully suppressed, this size-driven baseline brightness is expected to remain a substantial contributor to pixel loss, so operator-side glint avoidance alone would meaningfully reduce, but not eliminate, the impact characterized in this work. It is also only possible to implement for one satellite operator and one observatory at a time; it could inadvertently increase contamination at other astronomical facilities.

\textit{Twilight-avoiding pointing strategy.} Figure~\ref{fig:sky_density} shows the sunlit satellite population is concentrated in a specific, seasonally-varying region of sky near the terminator. Because this region is systematically avoidable by pointing choice, a scheduling strategy that preferentially points East during evening twilight and West during morning twilight -- rather than toward the terminator-adjacent region where illuminated satellites concentrate -- could substantially reduce twilight-time contamination. We do not attempt a full revised Rubin cadence simulation here, as this would require balancing this benefit against the resulting change to Rubin's own science footprint; we flag it as a promising scheduling-side mitigation meriting dedicated study, consistent with the sky-aware scheduling direction we raise in Section~\ref{sec:v5_comparison}.

\section{Impacts, Limitations, and Harm}\label{sec:science}

Each of the LSST science goals --- understanding the nature of dark matter and dark energy, creating an inventory of the Solar System, mapping the Milky Way, and exploring objects that change position or brightness over time --- would be differently yet significantly harmed by 100,000--1,000,000 ODC satellites in LEO and Sun-synchronous orbit as simulated here.

\subsection{Quantifying Science Impacts}
To quantify and ultimately mitigate the impact of satellite constellations on Rubin LSST science, it is necessary to consider four quantities: (1) the number of satellites and their orbital distribution during the decade of LSST, (2) the brightness distribution function of the satellite population, (3) how the Rubin optical system (including the Simonyi Survey Telescope and LSST Camera) respond to bright satellites passing through the field of view, and finally, (4) how images are processed and turned into catalogs of astrophysical sources and objects.
The present work defines a simulated ODC satellite constellation based on FCC filings for the first two quantities, namely, the number, orbital distribution, and intrinsic brightness of satellites.

Other works, such as \citet{tyson20,tyson24,snyder25,kandula26}, provide details on the LSST Camera and optical system and how it responds to bright linear features. This is the best constrained of the four quantities, and it is a strong function of wavelength, which is why Sections \ref{sec:streakSB} and \ref{sec:results_surface_brightness} take care to translate AB magnitudes based on monochromatic BRDFs into LSSTCam bandpass-aware surface brightnesses. The origin of the suggested radiant intensity limits presented in \citet{kandula26}, which are essentially a more detailed parametrization of the 7 V AB magnitude limit \citet{boley25} integrated over five of the six LSSTCam filter bands, is that the LSST camera exhibits non-linear electronic crosstalk between the 16 subsections of each CCD detector.

Finally, the Rubin Science Pipelines \citep{rsp} presently identify likely streaks and glints due to artificial objects in Earth orbit in two different stages of data analysis. For a more complete description of Rubin data products and data release processing steps, see \citet{dp1}. First, during deep stack (``coadd'') assembly, outliers and linear streak features are excluded. Second, during difference imaging, linear streak features in images and glint trails in detected source catalogs are flagged and removed from downstream object association. Work is ongoing to assess the efficacy of the difference imaging streak and glint detection, including improving the ability to detect faint streaks. The best outcome from image and catalog processing is an accurate indication of which data should be discarded or otherwise excluded from analysis. It is not possible to reconstruct information that would have appeared in an image in the absence of a satellite.

\subsection{Limitations}
It is important to note that such data losses are pernicious, cumulative, and can differ significantly depending on the science questions being asked and the subset of data being used. For instance, none of the four quantities adequately capture the changing sky brightness due to an increase in the scattered light background due to artificial objects in Earth orbit \citep[see, e.g.,][]{kocifaj21,hainaut26}. Our results indicate that the increase in sky brightness may even be the dominant science impact, but a more detailed investigation of this challenge is urgently needed. Neither do we consider the reasonable possibility that extremely large ODC satellites can transit astrophysical sources and affect observations even when they are not illuminated by the Sun.

We have further ignored the growing debris population, which will almost certainly continue to grow even more rapidly with the advent of ODCs. While much debris will remain below the LSST detection threshold, plenty of larger or shinier objects already manifest as trails of bright glints rather than wide streaks \citep{tyson24}. In addition, our simulated ODC satellite constellation affords us the luxury of knowing precisely the location of all satellites at any given time --- in reality, astronomers face significant challenges with regard to inaccurate ephemerides and distinguishing likely satellite signatures from real astrophysical phenomena in a way that minimizes false positives and also does not unintentionally discard rare discoveries.

\subsection{Observing Schedule as a Mitigation Factor: Comparison Between opsim v5.0 and v5.3}
\label{sec:v5_comparison}

The results presented in Sections~\ref{sec:results}--\ref{sec:mitigation}
are based on the v5.3 Rubin baseline scheduler simulation. To assess whether changes to the LSST observing strategy itself --- independent of satellite population size or altitude --- can meaningfully affect ODC contamination, we additionally repeat the full six-stage pipeline using the earlier v5.0 baseline scheduler simulation and compare the two. The v5.3 strategy differs from v5.0 primarily in survey bookkeeping rather than pointing philosophy: it incorporates a more realistic engineering and maintenance time model (46 weeks of annual maintenance versus 20), a Year-1 ``template tier'' that prioritizes acquiring low-seeing image pairs early in the survey, and an overall $\sim$10\% reduction in total visits relative to v5.0.

Table~\ref{tab:v5_comparison} summarizes the key contamination metrics under both scheduler versions. Several population-averaged quantities are more favorable under v5.3 than under v5.0: the 99th-percentile per-pointing pixel loss fraction is lower (47.26\% versus 57.42\%), the fraction of pointings suffering zero pixel loss is higher (82.2\% versus 79.64\%), and the fraction of pointings containing a detector-saturating streak is smaller (6.39\% versus 8.32\%). These differences appear to be an incidental consequence of the reduced total visit count in v5.3 rather than any satellite-aware adjustment to the scheduler itself.

At the same time, the fraction of all LSST observations occurring during twilight is \emph{higher} in v5.3 (25.1\%) than in v5.0 (22.47\%), driven by the new template-tier program's emphasis on early, paired, low-seeing acquisitions. Because twilight is precisely the regime in which ODC streaks are brightest and most numerous (Section~\ref{sec:seasonal}, Figure~\ref{fig:brightness_heatmap}), this shift works against the population-level gains described above: the single-pointing maximum pixel loss at 10\% ODC deployment is higher under v5.3 (87.25\%) than under v5.0 (81.5\%), and the fraction of pointings struck by an ultra-bright ($m_b < 2$) streak is similarly slightly higher (0.68\% versus 0.38\%).

This comparison suggests that survey-level scheduling choices can shift ODC contamination outcomes by several percentage points without any change to the satellite population itself, but that the \emph{direction} of the effect depends on when and where visits fall relative to twilight and the seasonally varying illuminated-satellite bands identified in Figure~\ref{fig:sky_density}. Simple scheduling changes, such as observing less at twilight or avoiding sky regions
where Sun-synchronous shells stay lit (Section~\ref{sec:sso_only}), could help retain v5.3's gains without its twilight trade-off. We do not attempt to quantify such a ``sky-aware'' scheduling algorithm here, but flag it as a promising, non-satellite-side mitigation avenue for future work, distinct from the population-, orbit-, and altitude-based strategies considered in Sections~\ref{sec:fewer}--\ref{sec:lower_altitude}.

\begin{deluxetable}{lcccc}
\tablecaption{Comparison of Key Contamination Metrics Between opsim
v5.0 and v5.3, at Full and 10\% Deployment \label{tab:v5_comparison}}
\tablewidth{0pt}
\tablehead{
\colhead{Metric} & \colhead{v5.0 (100\%)} & \colhead{v5.3 (100\%)} &
\colhead{v5.0 (10\%)} & \colhead{v5.3 (10\%)}
}
\startdata
Mean geometric crossings / night              & $\sim$218{,}000 & $\sim$180{,}000 & $\sim$21,756 & $\sim$18,134 \\
Mean illuminated streaks / night               & $\sim$88{,}000  & $\sim$60{,}000  & $\sim$8,799 & $\sim$6,026 \\
Sunlit streaks $\leq$ mag 12.5 (year total)    & $\sim$$1.3\times10^7$ & $\sim$$9\times10^6$   & $\sim$$1.3\times10^6$ & $\sim$$8.9\times10^5$ \\
Fraction brighter than mag 7 (IAU limit)       & 86.8\%          & 83.9\%          & 86.8\% & 83.66\% \\
Streaks brighter than mag 2                    & 174{,}000       & 229{,}000 & 17,610 & 20,899 \\
99th-pctile pixel loss per pointing            & 57.42\%          & 47.26\%          & 31.33\% & 29.95\% \\
Pointings with zero pixel loss                 & 79.64\%          & 82.2\%          & 79.81\% & 82.36\% \\
Pointings with saturating streak               & 8.32\%          & 6.39\%          & 8.13\% & 6.24\% \\
Pointings hit by ultra-bright ($m_b<2$) streak & 0.38\%          & 0.68\%          & 0.35\% & 0.64\% \\
Twilight fraction of all observations          & 22.47\%          & 25.1\%          & 22.47\% & 25.1\% \\
Max single-pointing pixel loss                 & 97.85\%            & 96.30\%            & 81.5\% & 87.25\% \\
Mean twilight pixel loss / night               & 7.65\%            & 6.88\%            & 2.5\%  & 2.33\% \\
\enddata
\tablecomments{v5.3 full-deployment values are drawn from the
one-year, full-deployment ($\sim$1M satellite) simulations described
in Section~\ref{sec:constellation} and presented throughout this
paper. v5.0 full-deployment values are from an equivalent run of the
same pipeline against the earlier v5.0 baseline scheduler simulation,
computed solely for this comparison. The 10\% deployment columns
refer to the reduced-population (1/10th-scale) scenario described in
Section~\ref{sec:fewer} and Appendix~\ref{app:tenth}, computed
separately for each opsim version.}
\end{deluxetable}

\subsection{Harm to Science}\label{sec:harm}
Ultimately, harm to science may be considered as an impact that is significant enough to either cause an action in response or to prevent an action that would have happened otherwise. Our simulated ODC constellation meets this threshold at both full deployment and at 10\% deployment (i.e., Section \ref{sec:fewer}). For example, discovering a new Solar System Object typically requires six separate observations \citep{kurlander2025predictions} , while a more distant transient event may only ever be seen once, depending on the characteristic timescale of the phenomenon. This can vary from, e.g., $\sim 1$ hour for stellar flares to $\sim 100$ days for certain supernovae. If a single source detection were rendered impossible due to a satellite streak, or an increase in background sky brightness, the new astrophysical object or phenomenon would not be discovered, and it would not be possible to schedule follow-up observations using complementary telescopes and facilities. Our results clearly demonstrate that a significant fraction of LSST pointings would contain a sunlit streak at all times of year and night (approximately $20\%$), and that many of these streaks would be considerably brighter than 7 AB magnitude (approximately $\mu_g = 13\,$mag arcsec$^{-2}$), rendering non-negligible quantities of image pixels useless for science and hindering discovery. This risk should be understood in the context of other, generally more frequent sources of missed or degraded observations, such as weather downtime: satellite-streak-driven loss of an irreproducible observation is one contributor among several competing sources of missed-observation risk, rather than the dominant one, and our intent here is to characterize the ODC-specific contribution to this risk rather than to claim it as the primary factor.

As another example, even relatively faint streaks that fall below the detection threshold when the defocus effect is fully accounted for --- fainter than approximately $\mu_g = 20\,$ mag arcsec$^{-2}$, which is rare for our simulated ODC constellation --- may still inhibit the ability of LSST to detect or accurately measure a faint source nearby. In this situation, a systematic error would be introduced, and it would be extremely challenging if not impossible to correct for. The harm is more challenging to quantify and much less obvious than the previous example. While faint streaks are still strongly preferable to bright ones, they too can harm LSST science.

Both of these examples simultaneously reinforce the conclusions of \citet{kandula26} and point to the recommendations therein being insufficient for the regime of our simulated ODC satellite constellation. Indeed, when their report was written, few of that report's coauthors thought that a satellite population exceeding 100,000 was a likelihood in the coming decade.

Our pixel loss analysis shows that during twilight pointings, which represent $\sim$25.07\% of all LSST observations, the mean pixel loss fraction across affected pointings is typically 5-10\% per night, with individual pointings occasionally losing more than 90\% of their science pixels. As described in Section~\ref{sec:pixelLoss}, this assumes a 100 px (20 arcsec) width for unsaturated streaks and a 500 px (100 arcsec) width for saturated streaks.

Averaged over the full survey year, approximately 6.39\% of all pointings contain at least one streak bright enough to saturate the detector (peak electron count exceeding the LSSTCam full-well capacity of $1.3\times10^5$\,e$^-$), directly rendering those pixels unrecoverable. A larger fraction of pointings contain unsaturated but still detectable streaks, which contribute additional pixel loss. Furthermore, approximately 0.68\% of all pointings are affected by at least one streak with $m_{b} < 2$\,mag, for which entire CCDs are lost due to bleed trails and charge-transfer artifacts (Section~\ref{sec:pixelLoss}). For the faintest transient phenomena, those that may only ever be observed in a single LSST pointing, a bright satellite streak crossing the relevant detector region means the event is simply missed, with no possibility of recovery or follow-up. For survey-mode science such as Solar System object discovery, the cumulative effect of even modest per-pointing pixel loss compounds across the required six observations, substantially reducing the effective survey area available for linkage and confirmation.

Along the same lines, the expected substantial increase in sky brightness during twilight lasting well into the night will all but eliminate the possibility of detecting faint sources, which could affect twilight surveys aimed at discovering potentially hazardous Solar System Objects \citep{schwamb2023tuning, kurlander2025predictions}.

Our results show that ODC satellite constellations as currently envisioned will have a significant and detrimental impact on the performance of the Rubin Observatory LSST. To preserve science and our skies, significant mitigating actions above and beyond what we explore in Section \ref{sec:mitigation} are urgently required.

\section{Summary and Conclusions}\label{sec:summary}

This study sheds some light onto the potential impact of proposed ODC satellite constellations on the Rubin Observatory LSST. The modeled constellation consists of nearly one million satellites distributed across 94 orbital shells spanning altitudes from approximately 686--1002 km and including both low-inclination and Sun-synchronous orbital configurations.

To quantify the impact on LSST observations, we develop a six-stage simulation pipeline that first identifies potentially affected pointings, then propagates satellites using SGP4, evaluates solar illumination and observer-frame geometry, estimates streak brightness using a BRDF-based reflectance model evaluated at 532\,nm, converts this to band-aware surface brightness, and finally estimates pixel loss due to streaks.

Our results show that ODC constellations would produce widespread contamination of LSST observations. Depending on season and observing geometry, up to 80\% of LSST pointings would contain streaks of illuminated satellites. Sun-synchronous shells remain illuminated more frequently and therefore contribute disproportionately to the observable contamination. 
The ODC brightness distribution is characterized by a diffuse reflection component around AB magnitude 6 and a brighter specular component reaching AB magnitudes of 2--4 driven by reflections from the large solar arrays required for orbital data center power generation. The brightest events occur during the two hours after sunset and before sunrise, when satellites remain sunlit during the astronomical night.

A per-band streak surface brightness analysis reveals that satellite streaks span a wide range of surface brightnesses across all six LSSTCam bands, from $\mu\sim$9.5 to 24.5\,mag\,arcsec$^{-2}$, where saturation occurs around 16\,mag\,arcsec$^{-2}$ (depending on band). The brightest events are concentrated in the twilight windows and are driven primarily by specular reflections from satellite solar arrays. Across all bands, the majority of streaks are classified as faint, falling below the per-pixel detection threshold, which may introduce systematic errors when sky background levels are measured and modeled. Still, a significant fraction reach sufficient surface brightness to saturate the detector, with peak electron counts exceeding the LSSTCam full-well capacity of $1.3\times10^5$\,e$^-$. The $u$-band is least affected due to the combination of lower system throughput and a positive solar color correction ($\Delta_u^{\odot} = +1.43$\,mag), while the $r$, $i$, $z$, and $y$ bands show the highest contamination rates consistent with the negative solar color corrections in those bands. Twilight pointings bear the greatest burden: the mean pixel loss fraction across twilight pointings is typically 5--10\% per night, with individual pointings occasionally exceeding 90\% pixel loss due to a single ultra-bright satellite crossing.

Estimates regarding the expected increase in sky brightness confirm earlier studies that predict a significant deterioration of observation conditions around twilight. In particular, during New Moon, the sky background can increase more than 10-fold (3--4 magnitudes) in $u,g,r$ and $i$ LSSTCam bands during the hours after sunset and before sunrise.

We note that our analysis is limited to impacts on Rubin Observatory LSST, which may be generalized in some ways to most ground- and LEO-based astronomy. However, our analysis does not include other important domains such as radio astronomy, orbital debris and collision risks, concerns about radiative cooling and impacts to infrared/sub-mm astronomy, impacts to astronomy from orbit raise and de-orbit activities, and myriad concerns pertaining to end-of-life satellite disposal schemes.

Overall, we find that mega-scale ODC constellations would substantially impact Rubin Observatory operations --- a conclusion that likely extends to other wide-field astronomical surveys \citep{hainaut26} and observatories in LEO \citep{kruk23,borlaff25}.
Contamination from large and bright satellite populations presents a significant challenge for astronomical imaging and transient science that no known mitigation strategy can effectively remedy.


\software{Astropy \citep{astropy2013, astropy2018, astropy2022}, 
          Healpy and HEALpix\footnote{\url{http://healpix.sourceforge.net}} \citep{healpix2005, Zonca2019},
          Matplotlib \citep{Hunter:2007},
          Numpy \citep{harris2020array},
          rubin\_sim \citep{yoachim2022},
          Scipy \citep{2020SciPy-NMeth},
          Skyfield \citep{Rhodes2019}, 
          Shapely \citep{shapely2007}
          ODC Simulation Pipeline \citep{jangid2026odcpipeline}}

\begin{acknowledgments}
This work has been supported by the National Science Foundation (NSF) under grant numbers AST-2332735 and AST-2332736. Any opinions, findings, and conclusions or recommendations expressed in this material are those of the author(s) and do not necessarily reflect the views of the NSF. 

This material is based upon work supported in part by the National Science Foundation through Cooperative Agreements AST-1258333 and AST-2241526 and Cooperative Support Agreements AST-1202910 and AST-2211468 managed by the Association of Universities for Research in Astronomy (AURA), and the Department of Energy under Contract No. DE-AC02-76SF00515 with the SLAC National Accelerator Laboratory managed by Stanford University. Additional Rubin Observatory funding comes from private donations, grants to universities, and in-kind support from LSST-DA Institutional Members.

This research makes use of data and/or services provided by the International Astronomical Union Centre for the Protection of the Dark and Quiet Sky (IAU CPS) SatHub. Any opinions, findings, and conclusions or recommendations expressed in this material are those of the author(s) and do not necessarily reflect the views of the NSF NOIRLab, the SKAO, ESO, the IAU, or any host or member institution of the IAU CPS.

We acknowledge support from the DiRAC Institute in the Department of Astronomy at the University of Washington. The DiRAC Institute is supported through generous gifts from the Charles and Lisa Simonyi Fund for Arts and Sciences, Janet and Lloyd Frink, and the Washington Research Foundation.
This work is supported by NSF NOIRLab, which is managed by the Association of Universities for Research in Astronomy (AURA) under a cooperative agreement with the U.S. National Science Foundation.

Finally, we thank Wil O'Mullane and the Rubin Publication Board as well as two anonymous IAU CPS Member reviewers for suggestions that greatly improved the manuscript.
\end{acknowledgments}

\bibliography{ms}{}
\bibliographystyle{aasjournalv7}

\appendix

\section{Streak Classification by Saturation Status}
\label{app:streak_count}

\begin{figure*}[!ht]
    \centering
    \includegraphics[width=\textwidth]{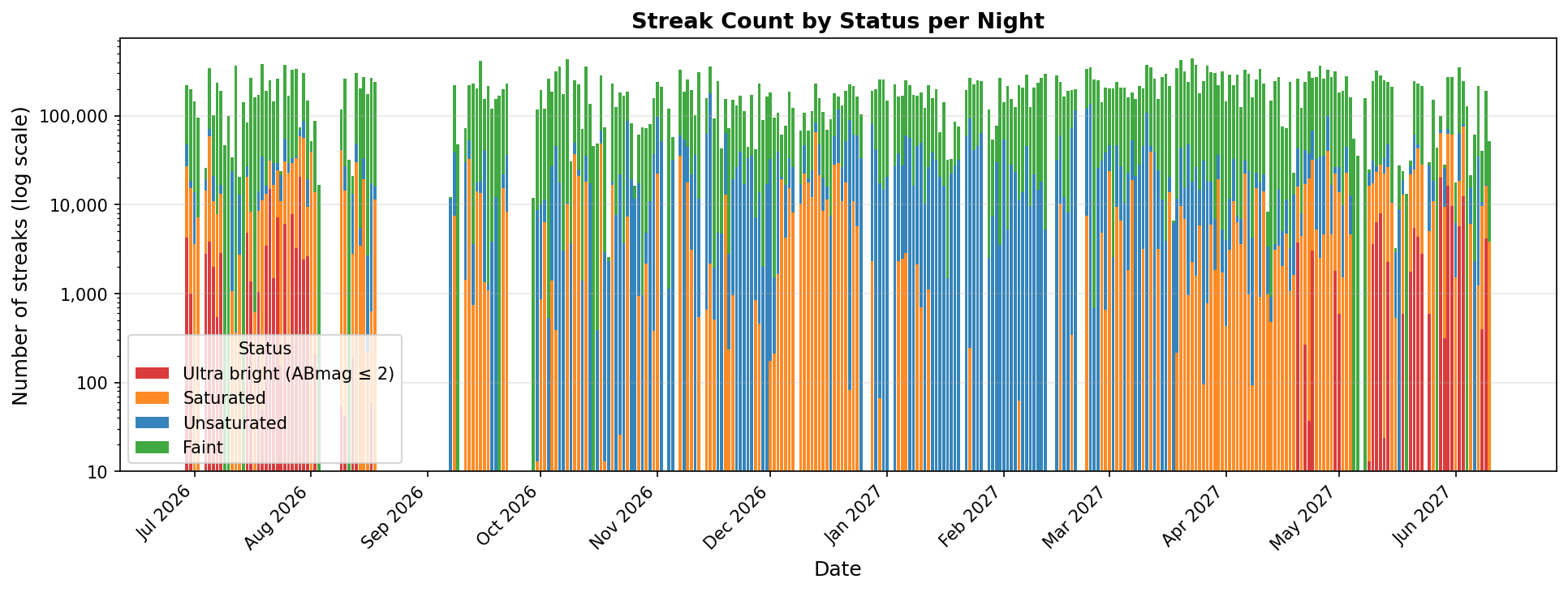}
    \caption{Number of satellite streaks per night classified by
    saturation status: Ultra bright(ABmag$<$2) (red), saturated (orange), unsaturated (blue), and
    faint (green), shown on a logarithmic scale. Gaps correspond to
    nights with no scheduled observations.}
    \label{fig:streak_count}
\end{figure*}

Figure~\ref{fig:streak_count} shows the number of satellite streaks classified by saturation status per night over the one-year survey period. Faint streaks (green) dominate throughout the year, comprising the vast majority of all crossings. Saturated (orange) and unsaturated (blue) streaks appear primarily during twilight periods when satellites are sunlit against a darker sky, and their counts rise and fall with the length of the twilight window over the seasonal cycle. There are also a few ultra-bright (red) streaks mainly in southern winter.

\section{Per-Band Streak Surface Brightness}
\label{app:sb_bands}

\begin{figure*}
    \centering
    \includegraphics[width=\textwidth]{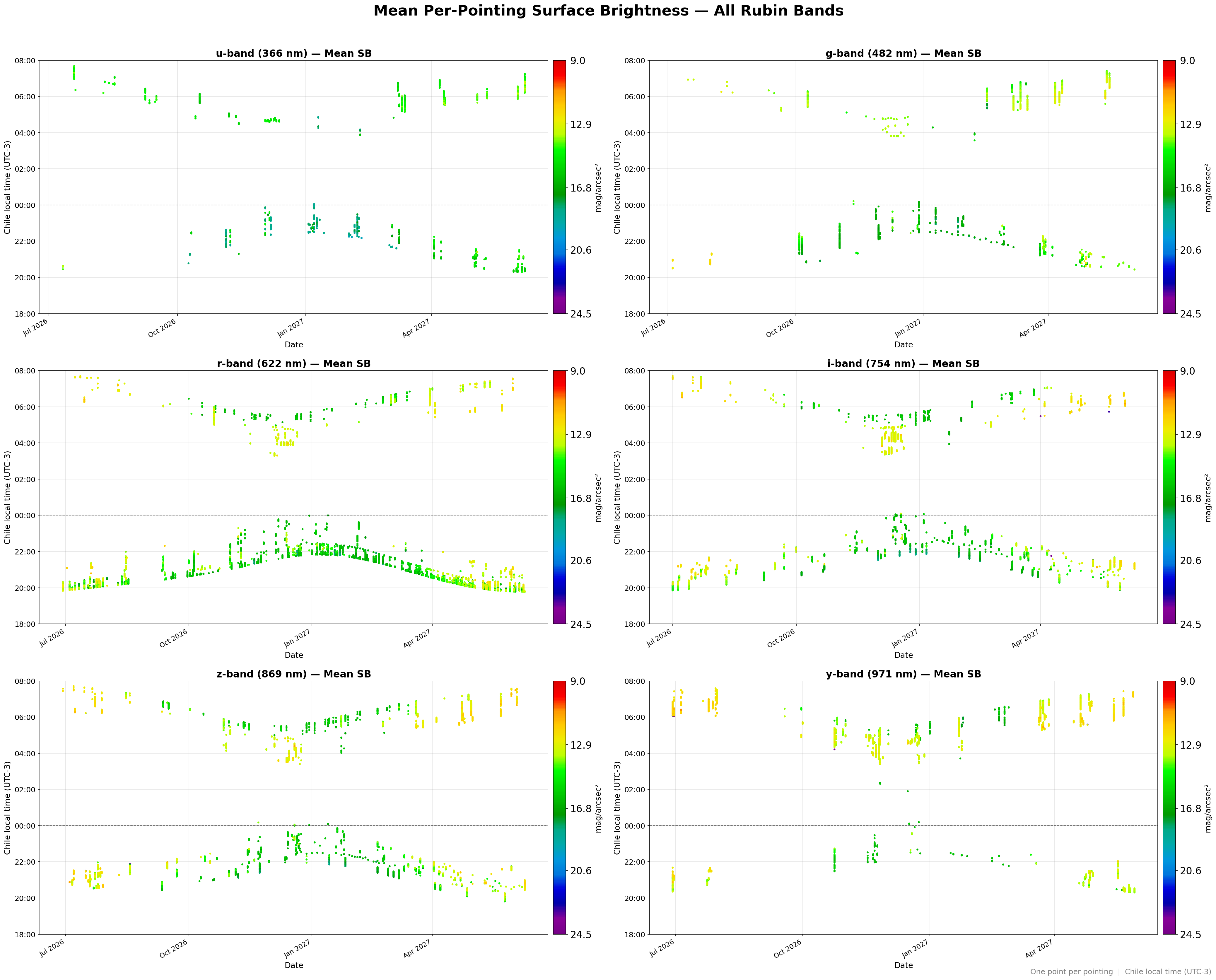}
    \caption{Mean streak surface brightness (mag\,arcsec$^{-2}$) in
    all six LSSTCam bands ($ugrizy$) as a function of date and Chile
    local time (UTC$-$3), per pointing. The dashed
    horizontal line marks local midnight. Brighter streaks (warmer
    colors, lower mag\,arcsec$^{-2}$) are concentrated in the
    twilight windows. The $u$-band shows markedly fewer streaks
    than the redder bands due to lower system throughput and the
    positive solar color correction at short wavelengths
    (Table~\ref{tab:bandparams}).}
    \label{fig:sb_bands}
\end{figure*}

Figure~\ref{fig:sb_bands} shows the mean streak surface brightness in each of the six LSSTCam bands ($ugrizy$) individually, as a function of date and Chile local time (UTC$-$3), per pointing. Several features are notable across the panels. First, the $u$-band (366\,nm) shows significantly fewer detected streaks than the redder bands, reflecting both the lower system throughput at short wavelengths and the positive solar color correction ($\Delta_u^{\odot} = +1.43$\,mag) which pushes satellite magnitudes fainter in $u$. Second, the $r$, $i$, $z$, and $y$ bands show progressively more streaks reaching brighter surface brightnesses (warmer colors), consistent with the negative solar color corrections in those bands. Third, in all bands the brightest streaks are concentrated in the twilight windows, where the combination of low satellite elevation and favorable solar illumination geometry produces the highest surface brightness events.

\section{Distribution of Satellite-Streak Pixel Loss}
\label{app:pixelLossDistribution}

\begin{figure}[ht]
\centering
\includegraphics[width=\textwidth]{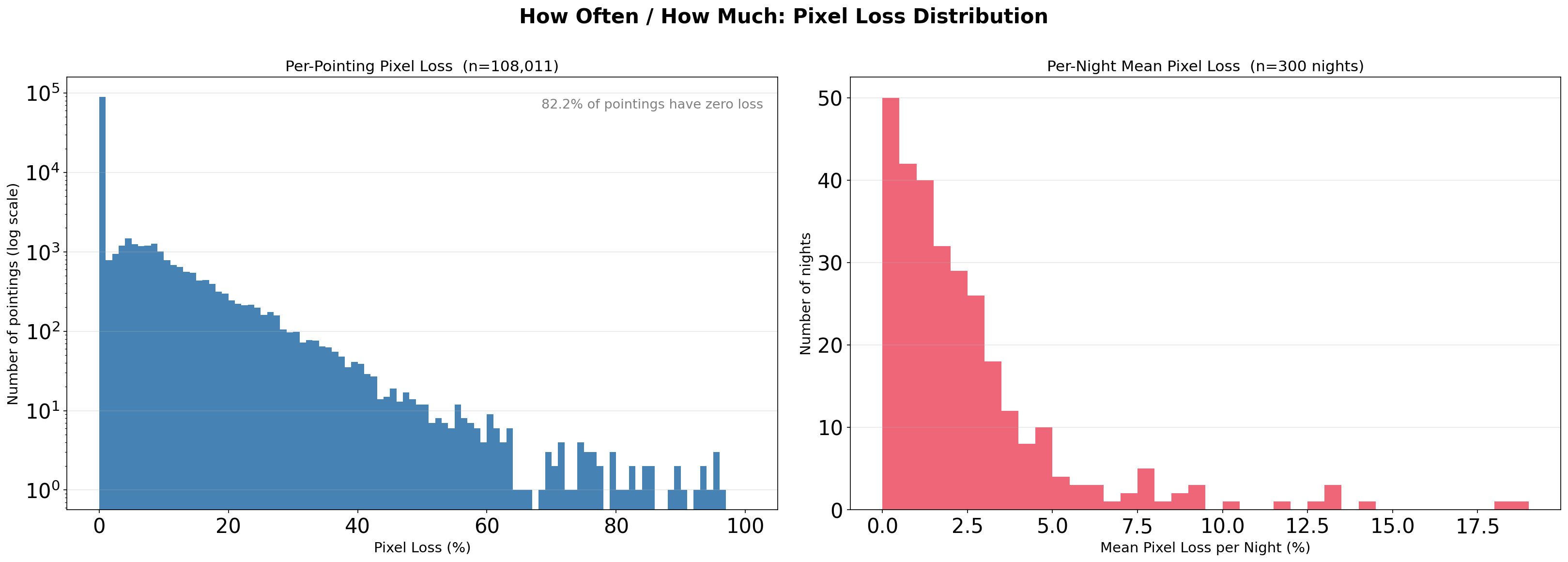}
\caption{
Left: Histogram of per-pointing pixel loss fraction due to satellite-streak masking, shown on a logarithmic count axis ($n=108{,}011$ pointings), drawn from the full night. The spike at $0\%$ indicates that $82.26\%$ of pointings are unaffected by streak masking. Right: Histogram of the mean pixel loss per night, averaged over all pointings taken that night ($n=300$ nights). The distribution is concentrated below $5\%$ but exhibits a tail extending to nights with mean losses near $20\%$.
}
\label{fig:pixelLossDistribution}
\end{figure}

\begin{figure}[ht]
\centering
\includegraphics[width=\textwidth]{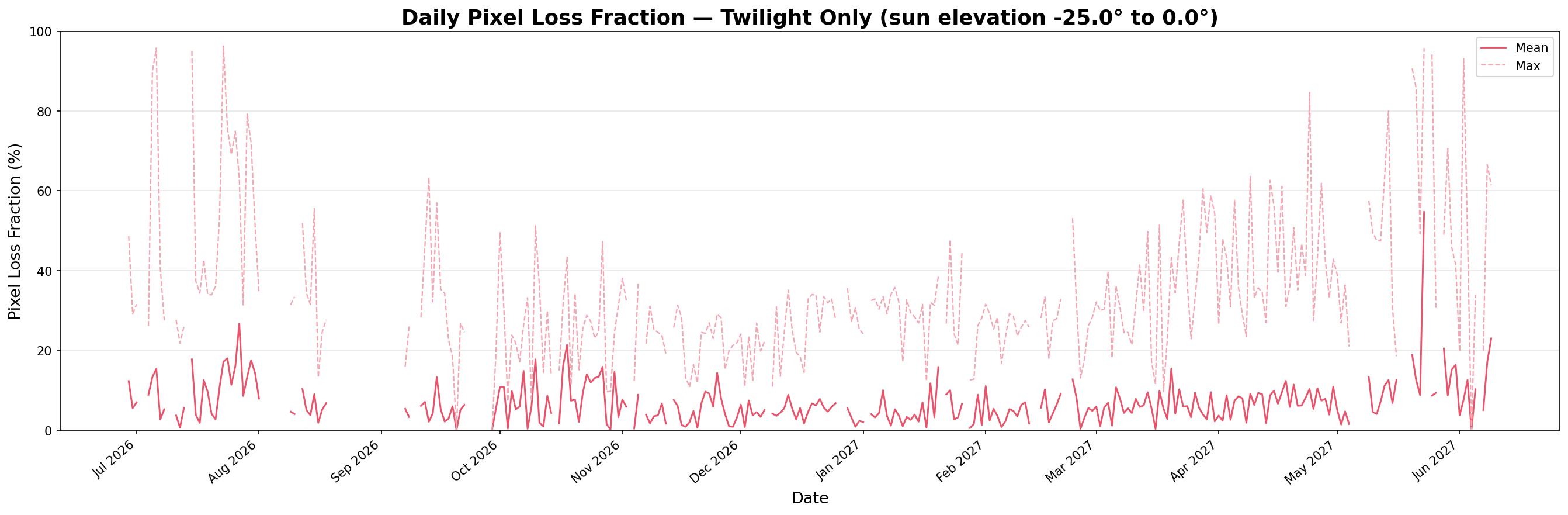}
\caption{
Daily recommended-mask pixel loss fraction for twilight pointings only (Sun elevation $-25\degr$ to $0\degr$) over the full survey baseline, showing the nightly mean (solid) and nightly maximum (dashed) across all twilight pointings observed that night. Gaps in the curves correspond to nights with no qualifying twilight pointings.
}
\label{fig:pixelLossTwilight}
\end{figure}

To complement the temporal view of pixel loss shown in Figure~\ref{fig:sb_heatmap}, we summarize the overall distribution of streak-induced pixel loss in Figures~\ref{fig:pixelLossDistribution} and~\ref{fig:pixelLossTwilight}.

Figure~\ref{fig:pixelLossDistribution} shows the distribution of pixel loss across the full night (18:00--08:00 local time), not restricted to twilight. The left panel shows the distribution of pixel loss fraction across all individual pointings ($n = 108{,}011$), on a logarithmic count axis. The vast majority of pointings, $82.2\%$, suffer no pixel loss at all from satellite streaks, producing the prominent spike at $0\%$. Among the remaining pointings, the distribution falls off roughly exponentially with increasing loss fraction, with a long tail extending to pointings that lose more than $80\%$ of their pixels to streak masking. The right panel aggregates this information by night, showing the distribution of the mean pixel loss per night across all pointings observed that night ($n = 300$ nights). Most nights have a mean pixel loss between $0\%$ and $5\%$, with the distribution's secondary peak around $2.5$--$3\%$. A small number of nights show substantially elevated mean pixel loss, up to nearly $20\%$, reflecting nights with an unusually high density of satellite crossings or particularly bright/extended streaks.

Figure~\ref{fig:pixelLossTwilight} restricts the same kind of analysis to twilight pointings only (Sun elevation $-25\degr$ to $0\degr$) and shows the night-by-night evolution over the full survey baseline. The solid curve gives the nightly mean recommended-mask pixel loss, while the dashed curve gives the nightly maximum. The nightly mean typically remains below $20\%$, with occasional spikes coinciding with periods of denser satellite traffic, most notably around August--September 2026, when the nightly mean briefly exceeds $30\%$ and the nightly maximum approaches $95\%$. The persistent gap between the mean and max curves throughout the baseline indicates that on most nights a small number of twilight pointings are much more strongly affected than the typical pointing, consistent with the long tail seen in the per-pointing histogram of Figure~\ref{fig:pixelLossDistribution}.

\begin{figure}[ht]
\centering
\includegraphics[width=\textwidth]{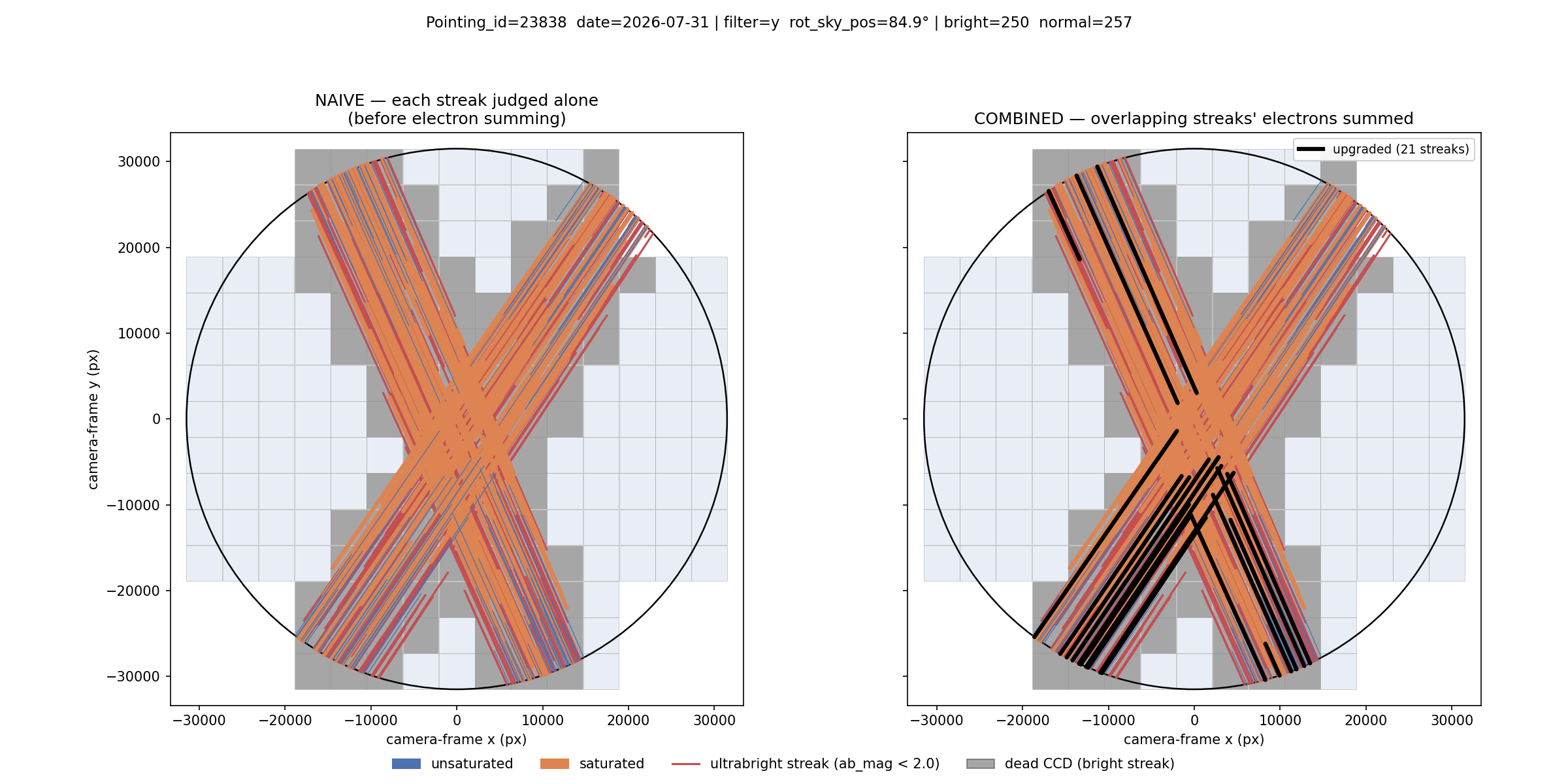}
\caption{
Fast (union-only, left) vs.\ slow (electron-summing, right) streak classification for pointing\_id~23838 (2026-07-31), a high-pixel-loss example pointing (total pixel loss $\approx 51.5\%$). Streak line width reflects the real mask width used in the pixel-loss calculation (100~px for unsaturated, 500~px for saturated streaks, Section~\ref{Streak Classification}); red lines are ultrabright ($m_b < 2$) streaks, each responsible for an entire dead CCD. In the right panel, 21 of 257 normal streaks (highlighted) are upgraded from unsaturated to saturated once overlapping streaks' electrons are summed -- despite being individually unsaturated when judged alone, as shown in the left panel.
}
\label{fig:naiveVsCombinedExample}
\end{figure}

To illustrate the electron-summing overlap mechanism underlying the slow-method pixel-loss values reported throughout this work (Section~\ref{Streak Footprint and Pixel Loss}), Figure~\ref{fig:naiveVsCombinedExample} shows a single high-pixel-loss example pointing in detail: pointing\_id~23838 (2026-07-31), with a total pixel loss fraction of approximately $51.5\%$. The left panel classifies each of this pointing's 257 normal streaks individually (the fast, union-only approach); the right panel reclassifies the same streaks after summing electrons from all overlapping streaks (the slow approach). Of the 257 normal streaks, 21 are upgraded from unsaturated to saturated once overlap is accounted for, despite being individually unsaturated when judged alone. Both panels also show this pointing's 250 ultrabright ($m_b < 2$) streaks in red, each of which renders its entire host CCD unusable independent of the electron-summing correction. This pointing was selected because its high total pixel loss reflects both mechanisms discussed in the main text: substantial dead-CCD loss from its large ultrabright population, and a dense overlapping-streak population in which the slow method measurably increases the saturated-streak count relative to the fast method.
 
\section{Increase in Night Sky Brightness}
\label{sec:appendix-sky-brightness}

In addition to the per-pointing pixel loss characterized in the main text, the ODC constellation contributes a diffuse increase to the overall night sky brightness at Rubin, integrated across every sunlit satellite simultaneously visible above the horizon at a given time, independent of whether any individual satellite crosses a specific pointing. For each timestep across an astronomical night (Sun elevation $\le -18^\circ$, extended through twilight to bracket the full observing window), we propagate every satellite in the simulated constellation, retain those that are both above the observing elevation cutoff and sunlit (Section~2.3), and sum their contributed flux at each of the six LSSTCam bands ($u$, $g$, $r$, $i$, $z$, $y$). This constellation-added flux is combined with the natural sky background to give
\begin{equation}
    \mu_{\rm combined} = -2.5 \log_{10}\!\left(f_{\rm natural} +
    \frac{F_{\rm total}}{\Omega}\right),
\end{equation}
where $F_{\rm total}$ is the summed constellation flux, $\Omega$ is the solid angle over which it is distributed, and $f_{\rm natural}$ is the natural sky background flux at that timestep, band, and sky position. The natural sky baseline is taken directly from Rubin's own sky brightness model \citep{RTN-012}, which provides the real, time- and band-dependent natural sky brightness -- including natural variation driven by the Moon's phase and position -- rather than a single fixed assumed value. We report both the resulting magnitude difference, $\Delta m = \mu_{\rm natural} - \mu_{\rm combined}$, and the fractional flux increase, $100 \times (F_{\rm total}/\Omega) / f_{\rm natural}$, at each timestep.

\begin{figure}[htbp]
    \centering
    \includegraphics[width=\textwidth]{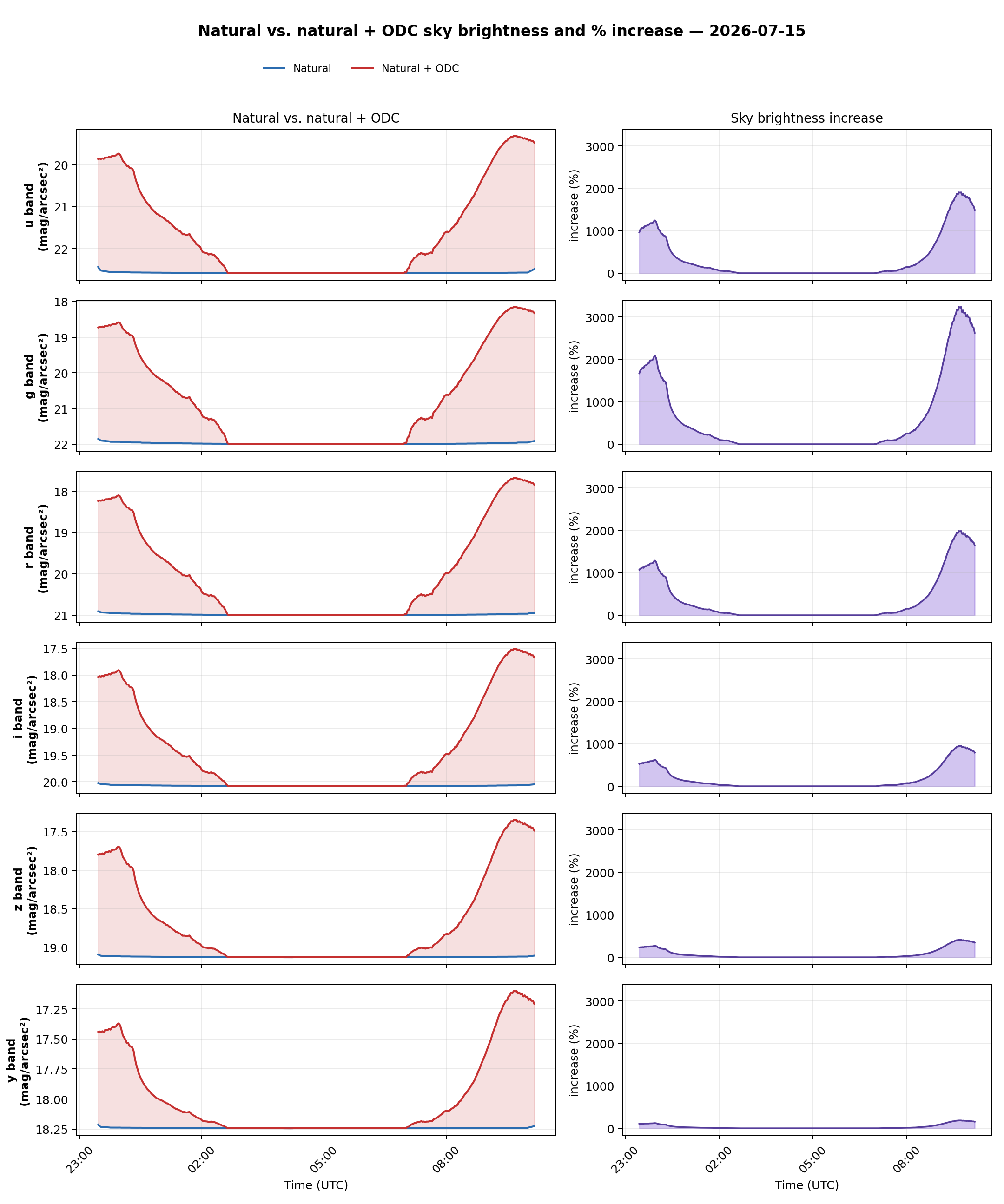}
    \caption{Left: Natural (blue) vs.\ natural~+~ODC (red) sky brightness increase for
     all six LSSTCam bands, for 2026-07-15, during New Moon
    (1.0\% illuminated Moon), Right: percent increase over the natural sky background.}
    \label{fig:sky-brightness-newmoon}
\end{figure}

\begin{figure}[htbp]
    \centering
    \includegraphics[width=\textwidth]{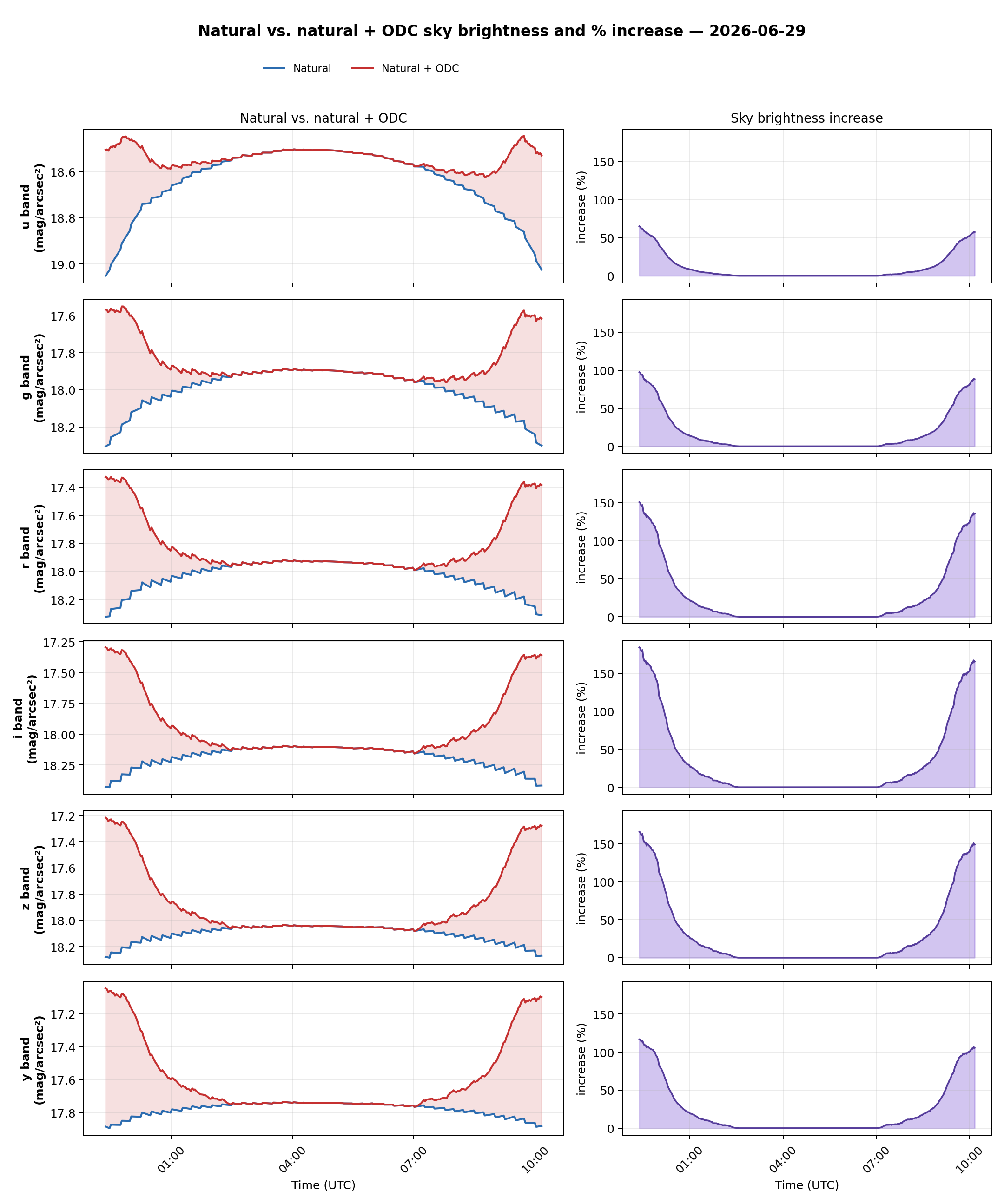}
    \caption{Same as Figure \ref{fig:sky-brightness-newmoon} but for 2026-06-29, during Full Moon (99.3\% illuminated Moon). }
    \label{fig:sky-brightness-fullmoon}
\end{figure}

Because the natural sky background itself varies by roughly an order of magnitude in flux between full moon and new moon, the \emph{same} constellation contribution produces a very different \emph{relative} impact depending on lunar phase. We illustrate this with two representative nights spanning the range of natural sky conditions in the simulated year: 2026-06-29 (99.3\% illuminated, effectively full moon) and 2026-07-15 (1.0\% illuminated, effectively new moon), shown in Figures~\ref{fig:sky-brightness-fullmoon}
and~\ref{fig:sky-brightness-newmoon} respectively, for all six bands.

 During New Moon nights (Figure~\ref{fig:sky-brightness-newmoon}) the increase in sky brightness during twilight hours is dramatic and can exceed 1000\% in $u$,$g$, $r$ and $i$ bands. Typically, $g$ and $r$ bands are used to discover potentially hazardous Solar System Objects during twilight. Only y-band observations can be expected to see somewhat less of an impact. In $u$, $g$ and $r$-bands the sky background is raised by up to three magnitudes well into the night.
 
 Full Moon nights have a higher natural background to begin with (Figure~\ref{fig:sky-brightness-fullmoon}), so that the bright, moon-dominated natural sky background dilutes the constellation's relative impact. However, the fractional increase in sky brightness still reaches approximately 150\% in every band across the night, with the bluer bands ($u$, $g$) being more affected than the redder bands ($z$, $y$).
 
 Both nights share the same qualitative structure: the fractional increase peaks sharply during evening and morning twilight, when the largest number of constellation satellites remain sunlit while the sky itself is still relatively dark, and drops to near zero during the deepest part of the night, when the great majority of satellites have entered Earth's shadow and no longer contribute reflected sunlight. This twilight concentration in the diffuse sky-brightness contribution mirrors the twilight concentration in per-pointing pixel loss discussed in Section~\ref{sec:results_pixel_loss}.

\section{ODC Constellation Sky Coverage and Pixel Loss at 10\% Deployment}
\label{app:tenth}

To support the scaling discussion in Section~\ref{sec:fewer}, we present sky-distribution and pixel-loss results computed for a 1/10th-scale realization of the full ODC constellation. Every 10th satellite per orbital plane per shell was retained, giving a self-consistent, evenly-sampled subset of 100,896 satellites across all 94 shells (versus $\sim$1~million at full deployment). Specifically, within each plane the satellites at indices $0, 10, 20, \ldots$ (by sorted name) were kept, ensuring that the subsampled constellation preserves the same true-anomaly phase coverage as the full shell rather than concentrating satellites at the start of each plane.

\subsection*{Sky Distribution}
 
\begin{figure*}[!ht]
    \centering
    \includegraphics[width=\textwidth]{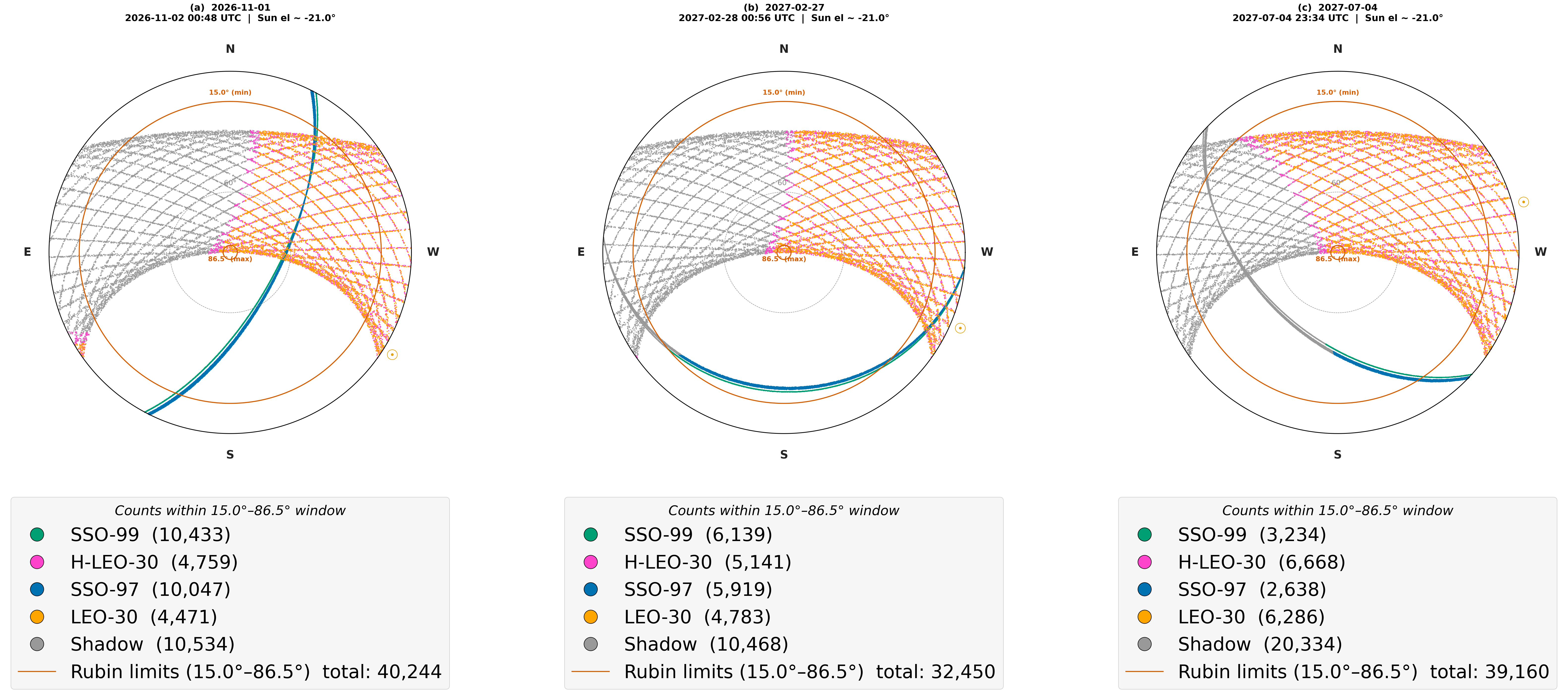}
    \caption{
        Stereographic sky maps of the ODC constellation at 1/10th deployment scale (100,896 satellites; every 10th satellite per orbital plane per shell retained) as seen from Rubin Observatory (lat $-30.24\degr$, lon $-70.75\degr$, elevation 2663\,m) at astronomical twilight (Sun elevation $\approx -21\degr$). Coloured points are sunlit satellites grouped by orbital regime: LEO-30 (orange), H-LEO-30 (magenta), SSO-97 (blue), SSO-99 (teal); see Table~\ref{tab:table1} for full shell definitions. Gray points are satellites in Earth's shadow. Solid orange circles mark the Rubin TMA operational elevation limits ($15\degr$ min, $86.5\degr$ max). Legend counts give the number of satellites within the Rubin observable window only.
    }
    \label{fig:sky_tenth}
\end{figure*}

Figure~\ref{fig:sky_tenth} shows stereographic sky maps of the 1/10th-scale constellation as seen from Rubin Observatory at astronomical twilight (Sun elevation $\approx -21\degr$) for three epochs spanning the survey year. Coloured points are sunlit satellites grouped by orbital regime (LEO-30, H-LEO-30, SSO-97, SSO-99; see Table~\ref{tab:table1}); gray points are satellites in Earth's shadow. The solid orange circles mark the Rubin TMA operational elevation limits ($15\degr$ minimum, $86.5\degr$ maximum). Counts in the legend refer to satellites within this observable window only.

All three panels reflect the same qualitative sky structure seen for the full constellation (Figure~\ref{fig:sky_density}), confirming that the 1/10th-scale subset faithfully reproduces the geometric and illumination properties of the full ODC population at reduced density.

\subsection*{Pixel Loss at 10\% Deployment}
 
\begin{figure*}[!ht]
    \centering
    \includegraphics[width=\textwidth]{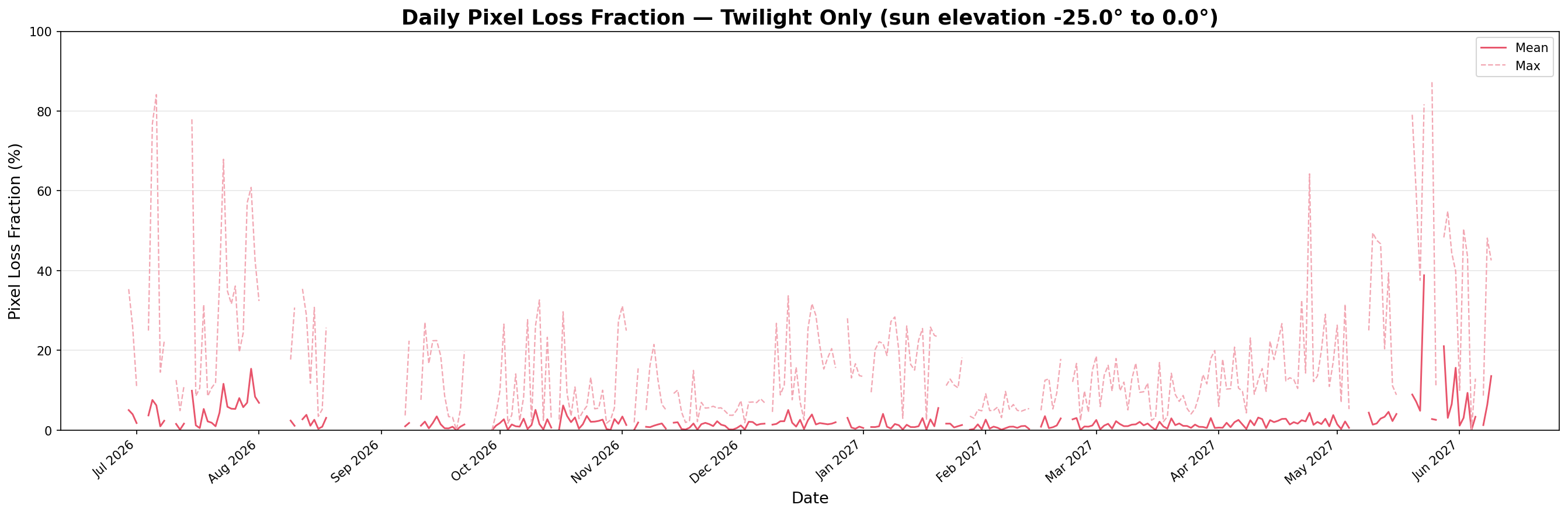}
    \caption{
        Daily pixel loss fraction for twilight pointings (Sun elevation $-25\degr$ to $0\degr$) computed from the 1/10th-scale ODC constellation simulation over the full one-year survey window, using the full pixel-loss pipeline (Section~\ref{sec:pixelLoss}). The solid line shows the mean pixel loss averaged over all twilight pointings on each night; the dashed line shows the maximum loss for any single pointing.
    }
    \label{fig:pixel_loss_tenth}
\end{figure*}

Figure~\ref{fig:pixel_loss_tenth} shows the daily pixel loss fraction for twilight pointings (Sun elevation $-25\degr$ to $0\degr$) computed from the 1/10th-scale simulation using the full pixel-loss pipeline described in Section~\ref{sec:pixelLoss}, including CCD-level kills for streaks with $m_{b} < 2$\,mag and polygon-union masks for fainter streaks.

Of 300 survey days, 283 contain at least one twilight pointing. At 1/10th scale, the mean pixel loss fraction across those days is $2.33\%$ and the single-pointing maximum reaches $87.25\%$. We note that pixel loss does not scale linearly with satellite number: because streak footprints within a pointing are union-ed before computing the loss fraction (Section~\ref{sec:pixelLoss}), even a 10-fold reduction in constellation size does not produce a 10-fold reduction in pixel loss. A single bright streak can still dominate an entire pointing regardless of how many other satellites are present. The seasonal pattern mirrors the full-constellation surface brightness result (Figure~\ref{fig:sb_heatmap}): the highest losses occur during the southern-hemisphere winter months (May--August), when longer twilight windows and favorable Sun--satellite--observer geometry increase both the number of illuminated satellites and the probability of bright specular reflection events.
 
These results confirm the conclusion drawn in Section~\ref{sec:fewer}: even at 10\% deployment, the ODC constellation produces substantial pixel loss throughout the survey year. The impact does not scale simply with satellite number because the brightest events are driven by specular reflection from large solar arrays --- a geometry-dependent effect that persists regardless of constellation density.
 
\section{Example Extremely Bright Satellite}\label{sec:superbright}

\begin{figure*}[!ht]
    \centering
    \includegraphics[width=\textwidth]{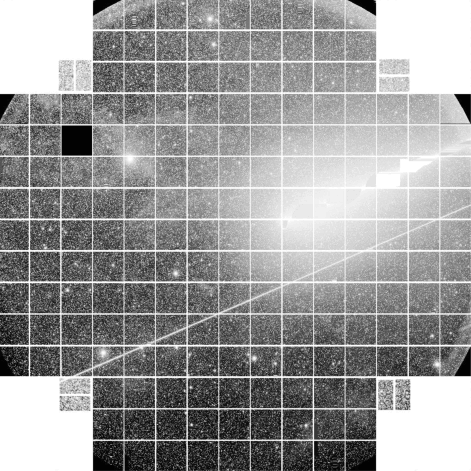}
    \caption{
        Two parallel satellite trails in a heavily binned mosaic version of an image taken by the LSST Camera in 2026 April. One streak is a more typical bright satellite that streaks across the full focal plane, and the other is overwhelmingly bright and stretches only partway across. There are clearly several detectors rendered entirely scientifically useless due to the extremely bright streak, and many more with severe systematic errors introduced to the background sky brightness. The image has had the initial steps of instrument signature removal performed (Post-ISR). Image credit: NSF-DOE Vera C. Rubin Observatory via RubinTV (included here with permission).
    }
    \label{fig:superbright}
\end{figure*}

Figure \ref{fig:superbright} shows a real Rubin image taken during the early operations period that happened to include a pair of bright satellite streaks. While streak photometry and identification is out of scope for this work, the brighter of the two streaks shown is clearly saturated, and several of the affected detectors failed to make it through the first few processing steps of the LSST Science Pipelines. Some comparable examples are described in \citet{polin26}. While overwhelmingly bright streaks like these are rare today, in an era of ODC constellations, this degree of contamination would be much more commonplace.

\end{document}